\documentclass[aps,prd,twocolumn,showpacs,groupedaddress,longbibliography,floatfix]{revtex4-1}
\pdfoutput=1
\usepackage{amsmath}
\usepackage{graphicx}
\usepackage{hyperref}
\usepackage{dcolumn}
\usepackage{longtable}
\usepackage{cancel}

\usepackage[normalem]{ulem}

\usepackage{color}
\definecolor{IITred}{rgb}{0.5,0.05,0.05}

\newcommand{\ket}[1]{| #1\rangle}

\newcommand{\euler}{\ensuremath{\gamma_{\text{E}}}}
\newcommand{\eq}[1]{Eq.\,(\ref{#1})}

\newcommand{\alphas}{\ensuremath{\alpha_{\mathrm{s}}}}
\newcommand{\etal}{\emph{et al.}}
\newcommand{\BR}{\ensuremath{\mathcal B}}
\newcommand{\T}{\rule{0pt}{2.6ex}}
\newcommand{\B}{\rule[-1.2ex]{0pt}{0pt}} 

\def\slashi#1{\rlap{\sl/}#1}

\newcommand{\mev}{\ensuremath{\hbox{ MeV}}}
\newcommand{\gev}{\ensuremath{\hbox{ GeV}}}
\newcommand{\tev}{\ensuremath{\hbox{ TeV}}}
\newcommand{\ps}{\ensuremath{\hbox{ ps}}}
\newcommand{\nb}{\ensuremath{\hbox{ nb}}}
\newcommand{\fb}{\ensuremath{\hbox{ fb}}}

\newcommand{\jpsi}{\ensuremath{J\!/\!\psi}}
\newcommand{\fm}{\ensuremath{\hbox{ fm}}}
\newcommand{\spec}[4]{\ensuremath{#1^{#2}\!{#3}_{#4}}}
\newcommand{\orcid}[1]{\thanks{\href{http://orcid.org/#1}{ORCID: #1}}}
\newcommand{\casefrac}[2]{{\textstyle \frac{#1}{#2}}}

\newcommand{\gammiss}{\ensuremath{\slashi{\gamma}}}
\begin{document}
\preprint{FERMILAB--PUB--16/nnn--T}

\title{Mesons with Beauty and Charm: \\ New Horizons in Spectroscopy} 


\author{Estia J. Eichten}
\email[]{Electronic mail: eichten@fnal.gov}
\orcid{0000-0003-0532-2300}
\author{Chris Quigg}
\email[]{Electronic mail: quigg@fnal.gov}
\orcid{0000-0002-2728-2445}
\affiliation{Fermi National Accelerator Laboratory \\ P.O. Box 500, Batavia, Illinois 60510 USA}


\date{\today} 

\begin{abstract}
The $B_c ^+$ family of $(c\bar{b})$ mesons with beauty and charm is of special interest among heavy quarkonium systems. The $B_c ^+$ mesons are intermediate between $(c\bar{c})$ and $(b\bar{b})$ states both in mass and size, so many features of the $(c\bar{b})$ spectrum can be inferred from what we know of the charmonium and bottomonium systems. The unequal quark masses mean that the dynamics may be richer than a simple interpolation would imply, in part because the charmed quark moves faster in $B_c$  than in the \jpsi. Close examination of the $B_c ^+$ spectrum can test our  understanding of the interactions between heavy quarks and antiquarks and may reveal where approximations break down.

Whereas the \jpsi\ and $\Upsilon$ levels that lie below flavor threshold are {metastable} with respect to strong decays, the $B_c$ ground state is absolutely stable against strong or electromagnetic decays. Its dominant weak decays arise from $\bar{b} \to \bar{c} W^{\star+}$, $c \to s W^{\star+}$, and $c \bar{b} \to W^{\star+}$ transitions, where $W^\star$ designates a virtual weak boson. Prominent examples of the first category are quarkonium transmutations such as $B_c^+ \to \jpsi\, \pi^+$ and $B_c^+ \to \jpsi\,\ell^+\nu_\ell$, where $\jpsi$ designates the $(c\bar{c})$ $1S$ level. 

The high data rates and extraordinarily capable detectors at the Large Hadron Collider give renewed impetus to the study of mesons with beauty and charm. Motivated by the recent experimental searches for the radially excited  $B_c$ states, we update the expectations for the low-lying spectrum of the $B_c$ system.  We make use of lattice QCD results, a novel treatment of spin splittings, and an improved quarkonium potential to obtain detailed predictions for masses and decays.  We suggest promising modes in which to observe  excited states at the LHC.  The $3P{\rm ~and~}3S$ states, which lie close to or just above the threshold for strong decays,  may provide new insights into the mixing between quarkonium bound  states and nearby two-body open-flavor channels. Searches in the $B^{(*)}\!D^{(*)}$ final states could well reveal narrow resonances in the $J^P=0^-, 1^-, {\rm ~and~} 2^+$ channels and possibly in the $J^P = 0^+{\rm ~and~}1^+$ channels at threshold.

Looking further ahead, the prospect of very-high-luminosity $e^+e^-$ colliders capable of producing tera-$Z$ samples  raises the possibility of investigating $B_c$ spectroscopy and rare decays in a controlled environment.   
\end{abstract}

\pacs{14.40.Lb, 14.40.Nd, 14.40.Pq\hfill \textsf{FERMILAB--PUB--19/075--T}}

\maketitle
\section{Introduction \label{sec:intro}}
Although the lowest-lying $(c\bar{b})$ meson has long been established, the spectrum of excited states is little explored. The ATLAS  experiment at CERN's Large Hadron Collider reported the observation of a radially excited 
$B_c$ state~\cite{Aad:2014laa}, but this sighting was not confirmed by  the LHC$b$ experiment~\cite{Aaij:2017lpg}.  The unsettled experimental situation and the large data sets now available for analysis make it timely for us to provide up-to-date theoretical expectations for the spectrum and decay patterns of narrow $(c\bar{b})$ states, and for their production in hadron colliders\,\cite{[{For a recent assessment, see contributions to the }][{, especially }]Microworkshop,*Yangmicro,*Luchinskymicro,*Oldemanmicro,*Berezhnoymicro}.  New work from the CMS Collaboration~\cite{Sirunyan:2019osb} shows the way toward exploiting the potential of $(c\bar{b})$ spectroscopy.

\subsection{What we know of the $B_c$ mesons \label{subsec:known}}
The possibility of a  spectrum of narrow $B_c$ states was first suggested by Eichten and Feinberg \cite{Eichten:1980mw}.
Anticipating the copious production of $b$-quarks at Fermilab's Tevatron Collider and CERN's Large Electron--Positron Collider (LEP), we presented a comprehensive portrait of the spectroscopy of the $B_c$ meson and its long-lived excited states~\cite{Eichten:1994gt}, based on then-current knowledge of the interaction between heavy quarks derived from $(c\bar{c})$ and $(b\bar{b})$ bound states, within the framework of nonrelativistic quantum mechanics~\cite{[{For other work in a similar spirit, see }][{}]Kiselev:1994rc,*Fulcher:1998ka,*Godfrey:1985xj, *Godfrey:2004ya,  *Ebert:2002pp, *Berezhnoy:1997fp, *Soni:2017wvy}. Surveying four representative potentials, we characterized the mass of the $J^P = 0^-$ ground state as $M(B_c) \approx 6258 \pm 20 \mev$. A small number of $B_c$ candidates appeared in hadronic $Z^0$ decays at LEP. The CDF Collaboration observed the decay $B_c^\pm \to \jpsi\, \ell^\pm \nu$ in 1.8-TeV $\bar{p}p$ collisions at the Fermilab Tevatron~\cite{Abe:1998wi}, estimating the mass as $M(B_c) \approx 6400 \pm 411 \mev$. (The generic lepton $\ell$ represents an electron or muon.) Subsequent work by the CDF~\cite{Aaltonen:2007gv}, D0~\cite{Abazov:2008kv}, and LHC$b$~\cite{Aaij:2012dd,*Aaij:2013gia,*Aaij:2014asa} Collaborations has refined the mass to $M(B_c) = 6274.9  \pm 0.8\mev$~\cite{Tanabashi:2018oca}, with the most precise determinations coming from fully reconstructed final states such as $\jpsi\, \pi^+$.

Investigations based on the spacetime lattice formulation of QCD aim to provide \emph{ab initio} calculations that incorporate the full dynamical content of the theory of strong interactions. Before the nonleptonic $B_c$ decays had been observed, a first unquenched lattice QCD prediction, incorporating $2+1$ dynamical quark flavors $(u/d, s)$ found $M(B_c) = 6304 \pm 12 ^{+18}_{-0}\mev$~\cite{Allison:2004be}, where the first error bar represents statistical and systematic uncertainties and the second characterizes heavy-quark discretization effects. Calculations incorporating $2 + 1 + 1$ dynamical quark flavors $(u/d, s, c)$~\cite{Dowdall:2012ab} yield $M(\spec{1}{1}{S}{0})=6278 \pm 9\mev$, in impressive agreement with the measured $B_c$ mass, and predict $M(\spec{2}{1}{S}{0}) = 6894 \pm 19 \pm 8\mev$~\footnote{We use spectroscopic notation \spec{n}{2S+1}{L}{J}, where $n$ is the principal quantum number, $S$ is total spin, and $L = S, P, D, \ldots$ represents the angular momentum $0, 1, 2, \ldots$}. 

Three distinct elementary processes contribute to the decay of $B_c$: the individual decays  $\bar{b} \to \bar{c} W^{\star+}$ and $c \to s W^{\star+}$ of the two heavy constituents, and the annihilation $c\bar{b} \to W^{\star+}$ through a virtual $W$-boson. Several examples of the $\bar{b} \to \bar{c}$ transition have been observed, including the final states $\jpsi \ell^+\nu_\ell$, $\jpsi \pi^+$, $\jpsi K^+$, $\jpsi\pi^+\pi^+\pi^-$, $\jpsi\pi^+ K^+K^-$, $\jpsi\pi^+\pi^+\pi^+\pi^-\pi^-$, $\jpsi D_s^+$, $\jpsi D_s^{\star +}$, and $\jpsi\pi^+ \bar{p}p$; and $\psi{(2S)}\pi^+$. A single channel, $B_s^0\pi^+$, representing the $c \to s$ transition is known. The annihilation mechanism, which would lead to final states such as $\tau^+\nu_\tau$ and $\bar{p}p\pi^+$, has not yet been established. The observed lifetime, $\tau(B_c) = (0.507 \pm 0.009)\ps$~\cite{Tanabashi:2018oca}, is consistent with theoretical expectations~\cite{Beneke:1996xe, *Anisimov:1998uk, *Kiselev:2000pp, *Chang:2000ac, Brambilla:2004wf}.  Predictions for partial decay rates (or relative branching fractions) await experimental tests.  Some recent theoretical works explore the potential of rare $B_c$ decays~\cite{Ali:2016gdg,*Esposito:2013fma,*Wang:2007sxa,*Wang:2015rcz}.

Until recently, the only evidence reported for a $(c\bar{b})$ excited state was presented by the ATLAS Collaboration~\cite{Aad:2014laa} in $pp$ collisions at $7\hbox{ and }8\tev$, in samples of $4.9\hbox{ and }19.2\fb^{-1}$. They observed a new state at $6842 \pm 7\mev$ in the $M(B_c^{\pm}\pi^{+}\pi^{-})-M(B_c^{\pm})-2M(\pi^{\pm})$ mass difference, with $B_c^{\pm}$ detected in the 
$J/\psi\, \pi^{\pm}$ mode. The mass ($527\pm 7\mev$ above $\langle M({1S})\rangle$) and decay of this state are broadly in line with 
expectations for the second $s$-wave state, $B^{\pm}_c{(2S)}$. In addition to the nonrelativistic potential-model calculations cited above, the HPQCD Collaboration has presented preliminary results from a lattice calculation using $2+1+1$ dynamical fermion flavors and highly improved staggered quark correlators~\cite{Lytle:2018ugs}. They report $M(\spec{2}{1}{S}{0}) = 6892 \pm 41\mev$, which is $576.5 \pm 41\mev$ above $\langle M({1S})\rangle$). This result and the NRQCD prediction~\cite{Dowdall:2012ab} lie above the ATLAS report by one and two standard deviations, respectively. The significance of the discrepancy is limited for the moment by lattice uncertainties. A plausible interpretation has been that ATLAS might have observed the transition $B_c^*(2S) \to B_c^*(1S)\pi^+\pi^-$, missing the low-energy photon from the subsequent $B_c^* \to B_c \gamma$ decay, and that the signal is an unresolved combination of \spec{2}{3}{S}{1}\ and \spec{2}{1}{S}{0}\ peaks. A search by the LHC$b$ collaboration in $2\fb^{-1}$ of 8-TeV $pp$ data yielded no evidence for either $B_c(2S)$ state~\cite{Aaij:2017lpg}. As we prepared this article for publication, the CMS Collaboration provided striking evidence for both $B_c(2S)$ levels, in the form of well-separated peaks in the $B_c\pi^+\pi^-$ invariant mass distribution, closely matching the theoretical template~\cite{Sirunyan:2019osb}. We incorporate these new observations into the discussion that follows in \S\ref{subsec:dipicasc}.

\subsection{Analyzing the $(c\bar{b})$ bound states \label{subseq:analyzing}} 
The nonrelativistic potential picture, motivated by the asymptotic freedom of QCD~\cite{Appelquist:1974zd}, gave early insight into the nature of charmonium and generated a template for the spectrum of excited states~\cite{Appelquist:1974yr, *Eichten:1974af}. For more than four decades, it has served as a reliable guide to quarkonium spectroscopy, including the states lying near or just above flavor threshold for fission into two heavy-light mesons that are significantly influenced by coupled-channel effects~\cite{Eichten:1978tg, *Eichten:1979ms,Eichten:2004uh, *Eichten:2005ga}.

We view the nonrelativistic potential-model treatment as a steppingstone, not a final answer, however impressive its record of utility. Potential theory does not capture the full dynamics of the strong interaction, and while the standard coupled-channel treatment is built on a plausible physical picture, it is not derived from first principles. Moreover, relativistic effects may be more important for $(c\bar{b})$ than for $(c\bar{c})$. The $c$-quark moves faster in the $B_c$ meson than in the \jpsi, because it must balance the momentum of a more massive $b$-quark. One developing area of theoretical research has been to explore methods more robust than nonrelativistic quantum mechanics~\cite{Brambilla:2004wf,Brambilla:2010cs,Sumino:2016sxe}.

Nonperturbative calculations on a spacetime lattice in principle embody the full content of QCD. This approach is yielding increasingly precise predictions for the masses of $(c\bar{b})$ levels up through $B_c^{*\prime} (\spec{2}{3}{S}{1})$ state. It is not yet possible to extract reliable signals for higher-lying states from the lattice, so we rely on potential-model methods to construct a template for the $B_c$ spectrum through the \spec{4}{3}{S}{1}\ level. If experiments should uncover systematic deviations from the expectations we present, they may be taken as evidence of dynamical features absent from the nonrelativistic potential-model paradigm, including---of course---coupling to states above flavor threshold, which we neglect our calculations of the spectrum.

In the following \S\ref{sec:theory}, we develop the theoretical tools required to compute the $(c\bar{b})$ spectrum. In earlier work~\cite{Eichten:1994gt}, we examined the Cornell Coulomb-plus-linear potential~\cite{Eichten:1978tg}, a power-law potential~\cite{Martin:1980jx}, Richardson's QCD-inspired potential~\cite{Richardson:1978bt}, and a second QCD-inspired potential due to Buchmueller and Tye~\cite{Buchmuller:1980su}, which we took as our reference model. We used a perturbation-theory treatment of spin splittings.
Using insights from lattice QCD and higher-order perturbative calculations, we construct a new potential that differs in detail from those explored in earlier work. We also use lattice results and rich experimental information on the $(c\bar{c})$ and $(b\bar{b})$ spectra to refine the treatment of spin splittings. We present our expectations for  the spectrum of narrow states in Section~\ref{sec:spectrum}.  We consider decays of the narrow states in section~\ref{sec:decays}, updating the results we gave in Ref.\,\cite{Eichten:1994gt}. We compute differential and integrated cross sections for the narrow $B_c$ levels in proton--proton collisions at the Large Hadron Collider in \S\ref{sec:lhcprod}.   Putting all these elements together, we show how to unravel the \spec{2}{}{S}{}\ levels and explore how higher levels might be observed.  
Prospects for a future $e^+e^- \to \text{Tera-}Z$ machine appear in \S\ref{sec:teraz}. We draw some conclusions and look ahead in Section~\ref{sec:conclude}.

\section{Theoretical preliminaries \label{sec:theory}}
We take as our starting point a Coulomb-plus-linear potential (the ``Cornell potential''\cite{Eichten:1978tg, *Eichten:1979ms}),
\begin{equation}
	V(r)=-\frac{\kappa}{r} + \frac{r}{a^2}\;\;,
	\label{eq:cornellpot}
\end{equation}
where $\kappa \equiv 4\alphas/3 =0.52$ and $a =2.34\gev^{-1}$ were chosen to fit the quarkonium spectra. Analysis of the \jpsi\ and $\Upsilon$ families led to the choices
\begin{equation}
m_c = 1.84\gev \qquad m_b = 5.18\gev .
\label{eq:Cornellparams}
\end{equation}
This simple form has been modified to incorporate running of the strong coupling constant in Refs.~\cite{Richardson:1978bt,Buchmuller:1980su}, among others, using the perturbative-QCD evolution equation at leading order and beyond. 
 At distances relevant for confinement, perturbation theory ceases to be a reliable guide. It is now widely held, following Gribov~\cite{Gribov:1999ui}, that as a result of quantum screening \alphas\ approaches a critical, or frozen, value at long distances (low energy scales).  In a light $(q\bar{q})$ system, Gribov estimated 
\begin{equation}
\alphas \to \overline{\alpha}_{\textrm{s}} = \frac{3\pi}{4}\left(1 - \sqrt{2/3}\right) \approx 0.14\pi = 0.44.
\label{eq:alphacrit}
\end{equation}
We incorporate the spirit of this insight into a new version of the Coulomb-plus-linear form that we call the \emph{frozen-\alphas\ potential.}

\begin{figure}[tb]
 \includegraphics[width=\columnwidth]{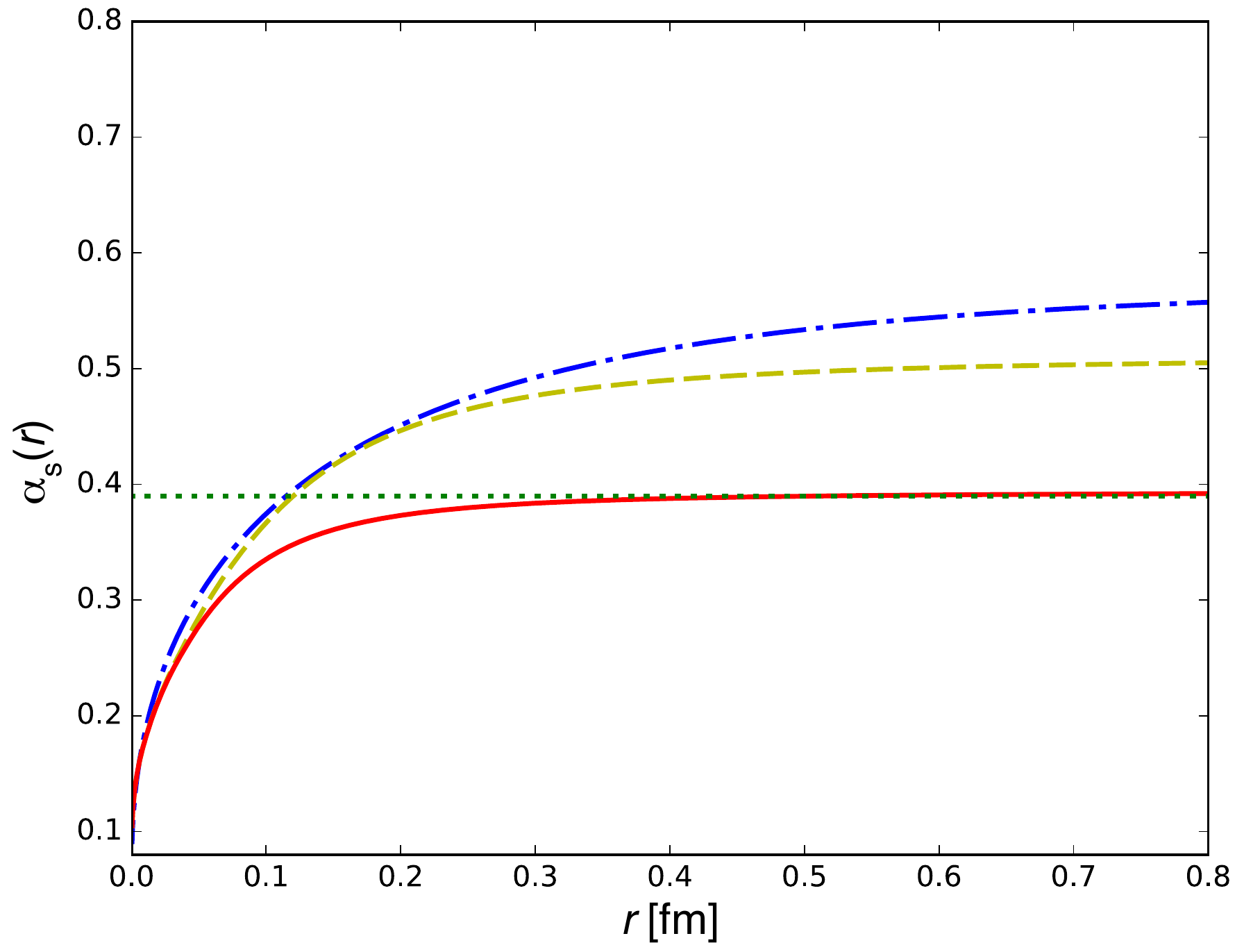}%
 \caption{Dependence of the running coupling $\alphas(r)$ on the interquark separation $r$. The strong coupling for our chosen potential is shown in the solid red curve. Those corresponding to the Cornell potential (green dots)~\cite{Eichten:1978tg}, Richardson potential (blue dash-dotted)~\cite{Richardson:1978bt} and an alternative version of the new potential with $\mu = 0.8\gev$ (gold dashes) are shown for comparison. \label{fig:alphar}}
 \end{figure}

The long-range part is the standard Cornell linear term. To obtain the Coulomb piece, we convert the four-loop running of $\alphas(q)$ in momentum space~\cite{Chetyrkin:2004mf,*Czakon:2004bu} to the behavior in position space using the method of~\cite{Jezabek:1998wk}, with an important modification. We set $\alphas(q = 1.6\gev) = 0.338$ and evolve with three active quark flavors. To enforce saturation of $\alphas(r)$ at long distances, we alter the recipe of Ref.~\cite{Jezabek:1998wk}, replacing the identification $ q = 1/r\exp(\euler)$, where $\euler = 0.57721 \ldots$ is Euler's constant, with the damped form $q = 1/[(r \exp(\euler)^2 + \mu^2]^{1/2}$. For our reference potential, we have chosen the damping parameter $\mu = 1.2\gev$. The consequent evolution of $\alphas(r)$ is plotted as the solid red curve in Figure~\ref{fig:alphar}, where we also show an alternative choice of $\mu = 0.8\gev$  (dashed gold curve), the constant \alphas\ of the original Cornell potential (dotted green curve) and $\alphas(r)$ corresponding to the Richardson potential (dot-dashed blue curve). 

\begin{figure}[tb]
 \includegraphics[width=\columnwidth]{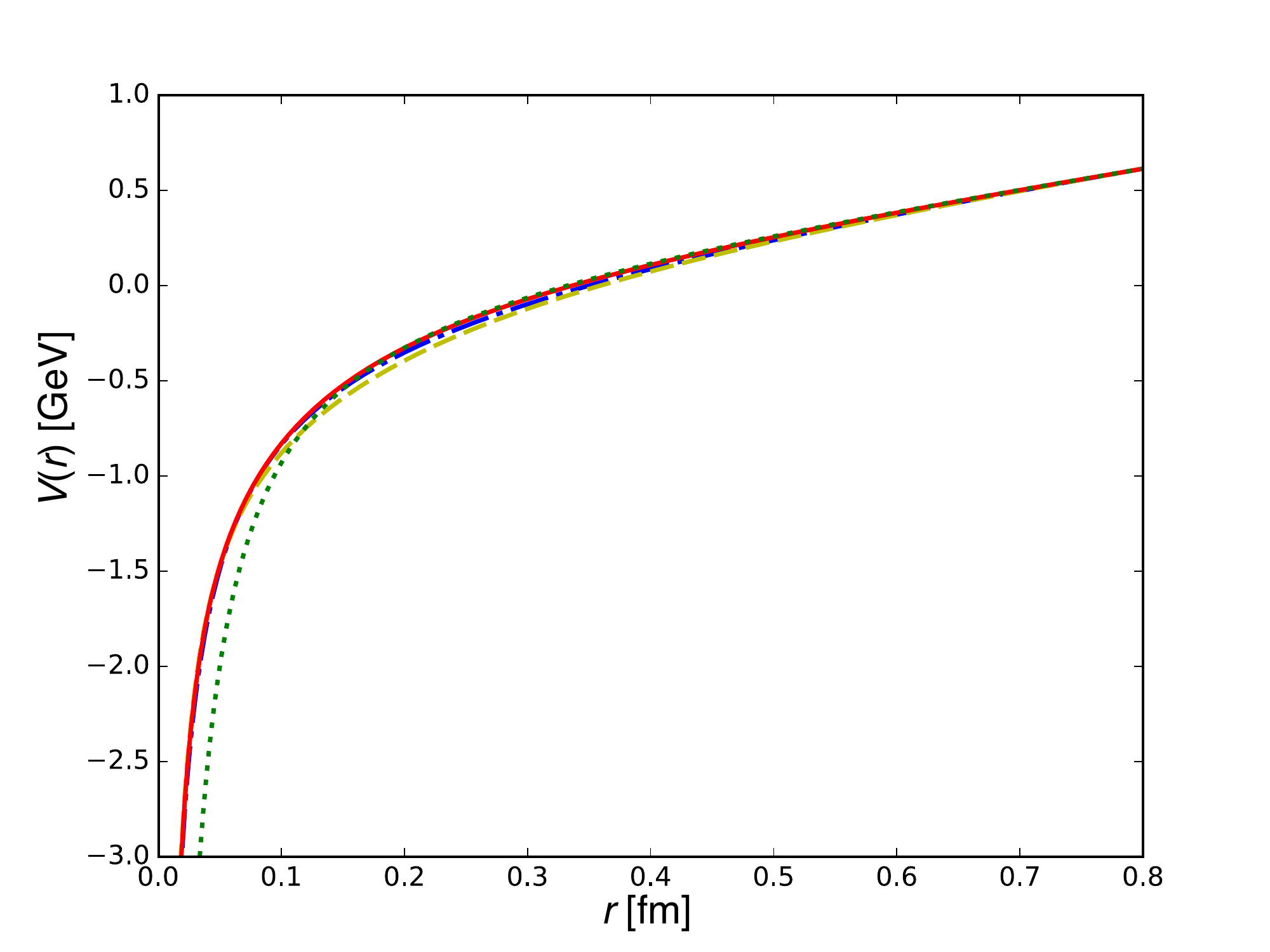}%
 \caption{Dependence of quarkonium potentials $V(r)$ on the interquark separation $r$. Our frozen-\alphas\ potential is shown in the solid red curve. The Cornell potential (green dots)~\cite{Eichten:1978tg}, Richardson potential (blue dash-dotted)~\cite{Richardson:1978bt}, and an alternative version of the new potential with $\mu = 0.8\gev$ (gold dashes) are shown for comparison. \label{fig:potcomp}}
 \end{figure}

We plot in Figure~\ref{fig:potcomp} the frozen-\alphas\ potential for both our chosen example, $\mu = 1.2\gev$, and the alternative, $\mu = 0.8\gev$. There we also show the Richardson and Cornell potentials. All coincide at large distances. The Cornell potential is deeper at short distances than any of the potentials that take account of the evolution of \alphas.
For the convenience of others who may wish to apply the new potential, we present values of $\alphas(r)$ suitable for interpolation in an Appendix. 

We presented the general formalism for spin-dependent interactions as laid out by Eichten \& Feinberg~\cite{Eichten:1980mw} and Gromes~\cite{Gromes:1984ma} in \S\,II\,B of Ref.~\cite{Eichten:1994gt}, where we took a perturbative approach to the spin--orbit and tensor interactions. In the intervening time, the charmonium and bottomonium spectra have been mapped in detail, as summarized in Table~\ref{tab:Psplits}. 
{\begingroup
 \squeezetable
\begin{table}[tb]
\caption{$P$-state masses~\cite{Tanabashi:2018oca} and splittings, in MeV.\label{tab:Psplits}}
 \begin{tabular}{cccc}
 \toprule
State & $2P\;(c\bar{c})$ &  $2P\;(b\bar{b})$ & $3P\;(b\bar{b})$ \T\\[2pt]
\colrule
$\chi_0(\spec{}{3}{P}{0})$ & $ 3414.71 \pm 0.30$  &  $9859.44 \pm 0.52$  & $10\,232.5 \pm 0.64$\T\\    
$\chi_1(\spec{}{3}{P}{1} )$&   $3510.67 \pm 0.05$  &  $9892.78 \pm 0.4$ & $10\,255.46 \pm 0.55$ \\ 
$h(\spec{}{1}{P}{1})$ &  $3525.38 \pm 0.11$  &  $9899.3 \pm 0.8 $ & $10\,259.8 \pm 1.12$ \\ 
$\chi_2(\spec{}{3}{P}{2})$ & $3556.17 \pm 0.07$  &  $9912.21 \pm 0.4$ &$10\,268.85 \pm 0.55$ \\[3pt] 
\spec{}{3}{P}{J}\ centroid, $\langle \chi \rangle$ & $3525.29 \pm 0.01$ & $9899.87 \pm 0.17$ & $10\,260.35 \pm 0.31$ \\[3pt]
$h - \langle \chi \rangle$ & $0.09 \pm 0.11$ & $-0.57 \pm 0.82$ & $-0.55 \pm 1.25$ \\
$\chi_1(\spec{}{3}{P}{1} ) -  \chi_0(\spec{}{3}{P}{0})$ &  $95.96 \pm 0.30$  &  $33.34 \pm 0.66$ & $22.96 \pm 0.84$ \\ 
$\chi_1(\spec{}{3}{P}{2} ) -  \chi_0(\spec{}{3}{P}{1})$ &  $45.5 \pm 0.09$  &  $19.43 \pm 0.57$ & $13.39 \pm 0.78$ \\[1pt]
\botrule
\end{tabular}
\end{table}
\endgroup}
This  wealth of information leads us now to choose a  more phenomenological approach. 

We write the spin-dependent contributions to the $(c\overline{b})$
masses as
\begin{equation}
\Delta = \sum_{k=1}^4 T_k \;\;\; ,
\end{equation}
where the individual terms are
\begin{eqnarray}
T_1 & = &
\frac{\langle\vec{L}\cdot\vec{s}_i\rangle}{2m_i^2}\widetilde{T}_1(m_i,m_j)
+ \frac{\langle\vec{L}\cdot\vec{s}_j\rangle}{2m_j^2}
\widetilde{T}_1(m_j,m_i) \nonumber \\
T_2 & = & \frac{\langle\vec{L}\cdot\vec{s}_i\rangle}{m_i m_j}
\widetilde{T}_2(m_i,m_j) + \frac{\langle\vec{L}\cdot\vec{s}_j\rangle}{m_i m_j}
\widetilde{T}_2(m_j,m_i)  \label{T1to4}\\
T_3 & = & \frac{\langle\vec{s}_i\cdot\vec{s}_j\rangle}{m_im_j}
\widetilde{T}_3(m_i,m_j) \nonumber \\
T_4 & = & \frac{\langle S_{ij}\rangle}{m_im_j} \widetilde{T}_4(m_i,m_j) \;\;,
\nonumber
\end{eqnarray}
$\vec{s}_i$ and $\vec{s}_j$ are the heavy-quark
spins, $\vec{S}=\vec{s}_i+\vec{s}_j$ is the total spin, $\vec{L}$ is the orbital angular momentum of
quark and antiquark in the bound state, $S_{ij} = 4\left[3(\vec{s}_i\cdot\hat{n})(\vec{s}_j\cdot\hat{n})-
\vec{s}_i\cdot\vec{s}_j\right]$ is the tensor operator, and $\hat{n}$ is an arbitrary unit vector.

We will deal with the hyperfine interaction $T_3$ momentarily. We express the other $\widetilde{T}_k$ as
\begin{eqnarray}
\widetilde{T}_1(m_i,m_j) & = & -\left\langle\frac{1}{r}\frac{dV}{dR}\right\rangle
   + 2\widetilde{T}_2(m_i,m_j) \nonumber \\
\widetilde{T}_2(m_i,m_j) & = & \frac{4\,\tilde{c}_2}{3}\left\langle \frac{\alphas(r)}{r^3} \right\rangle \label{eq:Ttildedefs}\\
\widetilde{T}_4(m_i,m_j) & = & \frac{\tilde{c}_4}{3}\left\langle \frac{\alphas(r)}{r^3} \right\rangle
\;,\nonumber
\end{eqnarray}
where we have introduced the phenomenological coefficients $\tilde{c}_2$ and $\tilde{c}_4$, which take the value unity in the perturbative approach.

We extract values of $\widetilde{T}_2$ and $\widetilde{T}_4$ for the observed levels that appear in Table~\ref{tab:Psplits}. These are shown as the underlined entries in Table~\ref{tab:coeff}. Then, we combine the definitions in \eq{eq:Ttildedefs} with our calculated values of $\langle \alphas/r^3\rangle$ to determine $\tilde{c}_2$ and $\tilde{c}_4$ in the $(c\bar{c})$ and $(b\bar{b})$ families. The geometric mean of these values is our estimate for the coefficients in the $c\bar{b}$ system. We insert these back into \eq{eq:Ttildedefs} to estimate the values of $\widetilde{T}_2$ and $\widetilde{T}_4$ for the $B_c$ family. For completeness, we include our evaluations of $\langle (1/r)dV/dr\rangle$ in the Table.
{\begingroup
 \squeezetable
\begin{table}[tbh]
   \centering
   \caption{Values of $\widetilde{T}_2$ and $\widetilde{T}_4$ extracted from data (\uline{underlined}) and the inferred values of the phenomenological coefficients $\tilde{c}_2$ and $\tilde{c}_4$ for the \jpsi\ and $\Upsilon$ families, from which the coefficients for the $(c\bar{b})$ system are derived.}
   \begin{tabular}{@{} lcccccc @{}} 
      \toprule
       System   &  $\widetilde{T}_2$ & $\widetilde{T}_4$ & $\left\langle\frac{\alphas(r)}{r^3}\right\rangle [\text{GeV}^{3}]$  
                   & $\tilde{c}_2$ & $\tilde{c}_4$ & $\left\langle\frac{1}{r}\frac{dV}{dr}\right\rangle [\text{GeV}^{3}]$ \T\B \\[3pt]
      \colrule
      $\spec{1}{}{P}{}(c\bar c)$ & $\uline{0.088}$ & $\uline{0.0308}$ & $0.0527$  & $1.25$ & $1.77$ & $0.141$ \T\\
      $\spec{1}{}{P}{}(b\bar b)$ & $\uline{0.258}$ & $\uline{0.0835}$ & $0.220$  & $0.82$ & $0.99$ & $0.383$\\
      $\spec{2}{}{P}{}(b\bar b)$ & $\uline{0.181}$ & $\uline{0.0547}$ & $0.166$  & $0.82$ & $0.99$ & $0.278$\\ 
      $\spec{1}{}{P}{}(c\bar b)$  & $0.119$ & $0.0388$ & $0.0885$   & $1.012$ & $1.316$ & $0.198$\B\\
         \botrule
   \end{tabular}
   \label{tab:coeff}
\end{table}
\endgroup}

For the \jpsi\ and $\Upsilon$ families, composed of equal-mass heavy quarks, the familiar $LS$ coupling scheme, in which states are labeled by \spec{n}{2S+1}{L}{J}, is apt. When the quark masses are unequal, as in the case at hand, spin-dependent terms in the Hamiltonian  mix the spin-singlet and spin-triplet $J=L$ states. We define
\begin{eqnarray}
\ket{\spec{n}{}{L}{L}}^\prime&  = & \cos\theta \ket{\spec{n}{1}{L}{L}} + \sin\theta\ket{\spec{n}{3}{L}{L}}  \nonumber \\[-3pt]
 & & \label{eq:angletheta} \\[-3pt]
\ket{\spec{n}{}{L}{L}}&  = & -\sin\theta \ket{\spec{n}{1}{L}{L}} + \cos\theta\ket{\spec{n}{3}{L}{L}} \; ,\nonumber
\end{eqnarray}
where 
\begin{equation}
 \tan{\theta} = \frac{2A}{B + \sqrt{B^2 +4A^2}},
 \label{eq:tandef}
 \end{equation}
 with
 \begin{equation}
A = \casefrac{1}{4} \sqrt{L (L+1)} \left(\frac{1}{m_c^2} -\frac{1}{m_b^2}\right)\widetilde{T}_1
\end{equation}
and 
\begin{equation}
B = \casefrac{1}{4} \left(\frac{1}{m_c^2} +\frac{1}{m_b^2}\right)\widetilde{T}_1 + \frac{1}{m_c m_b}\widetilde{T}_2  - 2\frac{1}{m_c m_b}\widetilde{T}_4 .
\end{equation}
Then our calculations of the $\widetilde{T}_k$ defined in \eq{eq:Ttildedefs} lead to these values for the mixing angle: $\theta_{2P}=18.7^\circ$, $\theta_{3D} = -49.2^\circ$, $\theta_{3P} = 21.2^\circ$, $\theta_{4F} = -49.5^\circ$, $\theta_{4D} = -40.3^\circ$. A Lattice calculation in quenched QCD~\cite{Davies:1996gi}  gave $\theta_{2P} = 33 \pm 2^\circ$. 

The masses of the mixed states are 
\begin{eqnarray}
M(\spec{n}{}{L}{L}^\prime)  &=& \langle M(\spec{n}{}{L}{})\rangle - \casefrac{1}{2}(B -\sqrt{4A^2+B^2}) \label{eq:mixmass} \\
M(\spec{n}{}{L}{L})   &=&  \langle M(\spec{n}{}{L}{})\rangle - \casefrac{1}{2}(B +\sqrt{4A^2+B^2}) , \nonumber
\end{eqnarray}
where$ \langle M(\spec{n}{}{L}{})\rangle$ is the \spec{n}{}{L}{} centroid.

At lowest order, the hyperfine splitting between $s$-wave states, arising from $T_3$, is given by
\begin{equation}
\Delta_{\text{HFS}}^{(n)} = M(\spec{n}{3}{S}{1}) - M(\spec{n}{1}{S}{0}) = \frac{8\alphas|R_{n0}(0)|^2}{9m_cm_b},
\label{eq:hfs}
\end{equation}
which is susceptible to significant quantum corrections. Rather than make \emph{a priori} calculations of the hyperfine splitting, we adopt the lattice QCD result for the ground state and scale the splittings of excited states according to
\begin{equation}
\frac{\Delta_{\text{HFS}}^{(n)}}{\Delta_{\text{HFS}}^{(1)}} = \frac{|R_{n0}(0)|^2}{|R_{10}(0)|^2}\;.
\label{eq:hfsrat}
\end{equation}

\section{The $B_c$ Spectrum  \label{sec:spectrum}}

The vector meson $B_c^*$, the \spec{1}{3}{S}{1} hyperfine partner of $B_c$ and analogue of \jpsi\ and $\Upsilon$, has not yet been observed. Modern lattice calculations~\cite{Gregory:2009hq, Dowdall:2012ab, Mathur:2018epb} give consistent values for the hyperfine splitting $M(B_c^*) - M(B_c)= (53\pm 7, 54\pm 7, 55\pm 3\mev)$, so we take the mass of the vector state to be $M(B_c^*) = 6329\mev$ and fix the centroid $\langle M({1S})\rangle$ of the ground-state $s$-wave doublet at $6315.5 \mev$ for the lattice.

We summarize in Table~\ref{tab:bcbarlevels} 
predictions for the spectrum of mesons with beauty and charm from our 1994 article~\cite{Eichten:1994gt}, lattice QCD calculations, and the present work, expressed as excitations with respect to the $1S$ centroid.   Other potential-model calculations, some incorporating relativistic effects, may be found in the works cited in Ref.\,\cite{Berezhnoy:1997fp}.
 {\begingroup
  \setlength\tabcolsep{2mm}
 \squeezetable
\begin{table}[tb]
\caption{Calculated excitation energies (in MeV) for $(c\bar{b})$ levels with respect to the $B_c({1S})$ centroid according to  potential models and Lattice QCD simulations. The potential models have been aligned with the $1S$ doublet centroid at $6315.5\mev$. Communication with states above flavor threshold is neglected.\label{tab:bcbarlevels}}
\begin{ruledtabular}
 \begin{tabular}{cccc}
Level & EQ94~\cite{Eichten:1994gt} &  Lattice QCD & \text{This Work}\\[2pt]
\hline 
\spec{1}{1}{S}{0} &  $-54.8$\phantom{M}    & $-40.5$~\cite{Gregory:2009hq, Dowdall:2012ab, Mathur:2018epb}\phantom{M} & $-40.5$\phantom{M}\T\\ 
\spec{1}{3}{S}{1} &  $18.2$    & $13.5$~\cite{Gregory:2009hq, Dowdall:2012ab, Mathur:2018epb} & $13.5$ \\ 
\hline
\spec{2}{3}{P}{0} & 381  & 393(17)(7) \cite{Mathur:2018epb} & $377$\T\\ 
\spec{2}{}{P}{1} & 411 &  417(18)(7) \cite{Mathur:2018epb} & $415$\\ 
$\spec{2}{}{P}{1}^\prime$ & 417  & 446(30) \cite{Davies:1996gi} & $423$ \\ 
\spec{2}{3}{P}{2} & 428  & 464(30) \cite{Davies:1996gi} & $435$\\ 
\spec{2}{1}{S}{0} & 537  & 561(18)(1) \cite{Dowdall:2012ab} & $551$\\ 
\spec{2}{3}{S}{1} & 580  & 601(19)(1) \cite{Dowdall:2012ab} & $582$\\ 
\hline
\spec{3}{3}{D}{1} & 693  &  -  & $691$\T\\ 
\spec{3}{}{D}{2} & 693 &  -  & $690$\\ 
$\spec{3}{}{D}{2}^\prime$ & 686 &  -  & $700$\\ 
\spec{3}{3}{D}{3} & 690  &  -  & $695$\\ 
\spec{3}{3}{P}{0} & 789  &  -  & $789$\\ 
\spec{3}{}{P}{1} & 823  &  -  & $820$\\ 
$\spec{3}{}{P}{1}^\prime$ & 816  &  -  & $828$\\ 
\spec{3}{3}{P}{2} & 834  &  -  &$839$ \\ 
\spec{3}{1}{S}{0} & 925  &  -  & $938$\\ 
\spec{3}{3}{S}{1} & 961  &  -  & $964$ \\
\hline
\spec{4}{3}{F}{2} & -  & - & $918$\T\\
\spec{4}{}{F}{3} & -  & - & $906$\\
$\spec{4}{}{F}{3}^\prime$ & -  & - & $922$\\
\spec{4}{3}{F}{4} & -  & - & $908$\\
\spec{4}{3}{D}{1} & -  & - & $1031$\\
\spec{4}{}{D}{2} & -  & - & $1033$\\
$\spec{4}{}{D}{2}^\prime$& -  & - & $1040$\\
\spec{4}{3}{D}{3} & -  & - & $1038$\\
\spec{4}{3}{P}{0} & -  & - & $1121$\\
\spec{4}{}{P}{1} & -  & - & $1149$\\
$\spec{4}{}{P}{1}^\prime$ & -  & - & $1158$\\
\spec{4}{3}{P}{2} & -  & - & $1167$\\
\spec{4}{1}{S}{0} & $1243$  & - & $1257$\\
\spec{4}{3}{S}{1} & $1276$  & - & $1280$\B\\
\end{tabular}
\end{ruledtabular}
\end{table}
\endgroup}

\begin{figure}[tb]
 \includegraphics[width=\columnwidth]{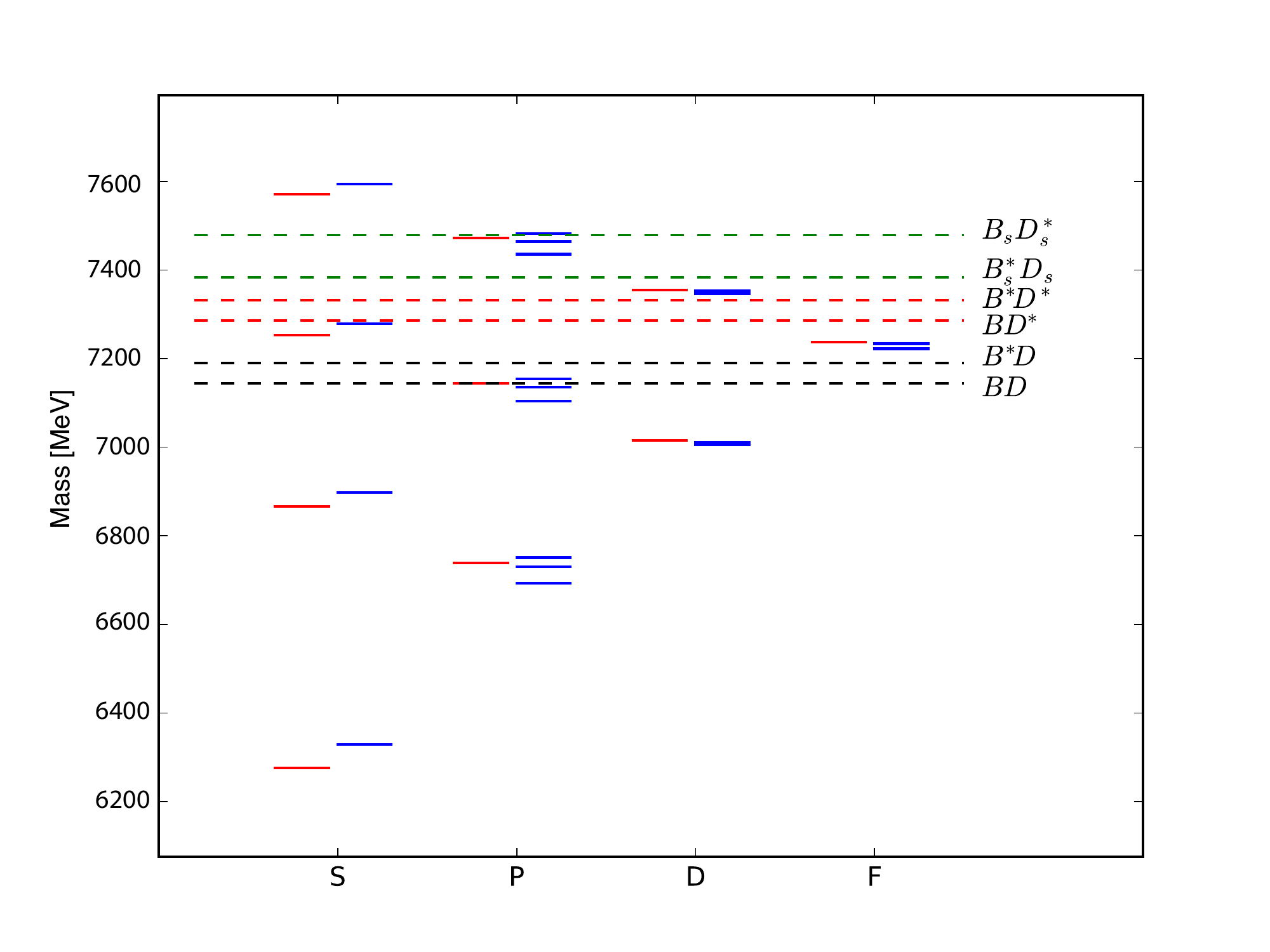}%
 \caption{Calculated $c\bar{b}$ spectrum, with (spin-singlet, spin-triplet) states shown on the (left [red], right [blue]) for each orbital-angular-momentum family S, P, D, F.  Dashed lines indicate thresholds for decay into two-body open-flavor channels given in Table~\ref{tab:bcbarthresh}.\label{fig:bcbarspec}}
 \end{figure}

Our expectations for the spectrum of states  are shown in the Grotrian diagram, Figure~\ref{fig:bcbarspec}, along with several of the lowest-lying open-flavor thresholds. 
 The thresholds for strong decays of excited $(c\bar{b})$ levels are known experimentally to high accuracy, as shown in Table~\ref{tab:bcbarthresh}.  
 {\squeezetable
 \begin{table}
 \caption{Open-flavor $(c\bar{b})$ thresholds and excitations above $1S$ centroid, $6315.5\mev$, for  $B_c$ levels, in MeV. \label{tab:bcbarthresh}}
 \begin{ruledtabular}
 \begin{tabular}{ccc}
 State & Flavor threshold & Excitation energy \\[2pt]
 \hline
  $B^+D^0$ & $7144.15 \pm 0.15$ & $829$\T\\
 ${B}^0{D}^+$ & $7149.28 \pm 0.16$ & $834$\B\\
 $B^{*+}D^0$ & $7189.48 \pm 0.26$ & $874$\\
${B}^{*0}{D}^+$ & $7194.30 \pm 0.26$ & $879$\B\\
$B^+D^{*0}$ & $7286.17 \pm 0.15$ & $971$ \\
$B^0D^{*+}$ & $7289.89 \pm 0.16$ & $974$\B\\
 $B^{*+}D^{*0}$ & $7331.50 \pm 0.26$ & $1016$\\
 $B^{*0}D^{*+}$ & $7334.91 \pm 0.26$ & $1019$\B\\
 ${B}^0_sD_s^+$ & $7335.23 \pm 0.20$ & $1020$\B\\
  ${B}_s^0D_s^{*+}$ & $7479.09 \pm 0.44$ & $1164$ \\
${B}_s^{*0}D_s^+$ & $7383.74^{+1.80}_{-1.50}$ & $1068$\B\\
 $B_s^{*0}D_s^{*+}$ & $7527.60^{+1.84}_{-1.55}$ & $1212$\\[1pt]
  \end{tabular}
 \end{ruledtabular}
 \end{table}
 }
 Comparing with the model calculations summarized in Table~\ref{tab:bcbarlevels}, we conclude that two sets of narrow $s$-wave $(c\bar{b})$ levels will lie below the beauty+charm flavor threshold, in agreement with general arguments~\cite{Quigg:1977xd}. All of the potential models cited in Ref.\,\cite{Ebert:2002pp} predict \spec{3}{3}{S}{1} masses well above the 829-MeV $B{D}$ threshold. For the \spec{3}{1}{S}{0} level, only the Ebert \etal\ prediction does not lie significantly above $B^*{D}$ threshold. Lattice QCD calculations do not yet exist for states beyond the $2S$ levels. 

\section{Decays of narrow $(c\bar{b})$ levels \label{sec:decays}}
 \subsection{Electromagnetic transitions \label{subsec:E1M1}}
 The only significant decay mode for the \spec{1}{3}{S}{1} $(B_c^*)$ state is the magnetic
dipole (spin-flip) transition to the ground state, $B_c$.  The M1 rate for transitions
between $s$-wave levels is given by
\begin{equation}
\Gamma_{\text{M1}}(i\rightarrow f+\gamma) = \frac{16\alpha}{3}\mu^2 k^3
(2J_f+1)|\langle f|j_0(kr/2)|i\rangle|^2\;\;\;,
\end{equation}
where the magnetic dipole moment is
\begin{equation}
\mu = \frac{m_be_c-m_ce_{\overline{b}}}{4m_cm_b}\;\;\;
\end{equation}
and $k$ is the photon energy.

 Apart from that M1 transition, only the electric dipole transitions are
important for mapping the $(c\bar{b})$ spectrum.
The strength of the electric-dipole transitions is governed by the size
of the radiator and the charges of the constituent quarks.  The E1
transition rate is given by
\begin{equation}
\Gamma_{\text{E1}}(i\rightarrow f+\gamma) = \frac{4\alpha \langle e_Q \rangle^2}{27}
k^3 (2J_f+1)|\langle f|r|i\rangle |^2 {\cal S}_{if}\;\;\; ,
\end{equation}
where the mean charge is
\begin{equation}
\langle e_Q \rangle = \frac{m_be_c-m_ce_{\overline{b}}}{m_b+m_c} \;\;\; ,
\end{equation}
$k$ is the photon energy, and the statistical factor ${\cal S}_{if}={\cal
S}_{fi}$ is as defined by Eichten and Gottfried~\cite{Eichten:1976jk}.  ${\cal
S}_{if}=1$ for $\spec{}{3}{S}{1}\rightarrow \spec{}{3}{P}{J}$ transitions and ${\cal
S}_{if}=3$ for allowed E1 transitions between spin-singlet states.  The
statistical factors for $d$-wave to $p$-wave transitions are reproduced in
Table~\ref{tab:stat}.
{
  \begin{table}[b!]
\caption{Statistical factor ${\cal S}_{if}={\cal S}_{fi}$ for
$^3P_J \rightarrow {^3D}_{J^\prime}+\gamma$  and $^3D_J \rightarrow {^3P}_{J^\prime}+\gamma$ transitions. }
\begin{tabular}{ccc}
\toprule
$J$ & $J^\prime$ & ${\cal S}_{if}$\B\T\\
\colrule
0 & 1 & 2 \\
1 & 1 & 1/2 \\
1 & 2 & 9/10 \\
2 & 1 & 1/50 \\
2 & 2 & 9/50 \\
2 & 3 & 18/25\B\\
\botrule
\end{tabular}
\label{tab:stat}
\end{table}
}

The significant M1 and E1 electromagnetic transition rates and the $\pi\pi$ cascade rates are given in Table~\ref{tab:totalwidth}, along with the total widths in the absence of strong decays.
\begin{widetext}
{\begingroup
\squeezetable
\begin{table*}[tbh] 
\caption{Total widths $\Gamma$ and branching fractions $\mathcal{B}$ for principal decay modes of $(c\bar{b})$ states below threshold, updating Table~IX of Ref.~\cite{Eichten:1994gt}.  Dissociation into $BD$, etc., will dominate over the tabulated decay modes for states  above  threshold. }
\begin{minipage}{0.45\textwidth} 
\begin{centering} 
\begin{ruledtabular}
\begin{tabular}{ lcc } 
Decay Mode & $k_{\gamma}$\,\text{[keV]} & Branching Fraction (\%) \B\\ 
\colrule 
\multicolumn{3}{c}{\spec{1}{1}{S}{0} (6275) : ~~~ weak decays} \T\\ 
\colrule 
\multicolumn{3}{c}{\spec{1}{3}{S}{1} (6329) : $\Gamma = 0.144\text{ keV}$} \T\\ 
$\spec{1}{1}{S}{0}+\gamma$ & $54$ & 100 \\ 
\colrule 
\multicolumn{3}{c}{\spec{2}{3}{P}{0} (6692) : $ \Gamma =53.1$ keV}  \T\\ 
\spec{1}{3}{S}{1} (6329) & 354 & $100$ \\ 
\colrule 
\multicolumn{3}{c}{\spec{2}{}{P}{1} (6730) : $ \Gamma =72.5$ keV} \T \\ 
\spec{1}{3}{S}{1} (6329) & 389 & $86.2$ \\ 
\spec{1}{1}{S}{0} (6275) & 440 & $13.7$ \\ 
\colrule 
\multicolumn{3}{c}{$\spec{2}{}{P}{1}^\prime$ (6738) : $ \Gamma =99.9$ keV} \T \\ 
\spec{1}{1}{S}{0} (6275) & 448 & $92.4$ \\ 
\spec{1}{3}{S}{1} (6329) & 397 & $7.51$ \\ 
\colrule 
\multicolumn{3}{c}{\spec{2}{3}{P}{2} (6750) : $ \Gamma =79.7$ keV}  \T\\ 
\spec{1}{3}{S}{1} (6329) & 409 & $100$ \\ 
\colrule 
\multicolumn{3}{c}{\spec{2}{1}{S}{0} (6866) : $ \Gamma =73.1$ keV} \T \\ 
$\spec{1}{1}{S}{0} + \pi \pi $ && $81.1$ \\ 
$\spec{2}{}{P}{1}^\prime$ (6738) & 126 & $16.5$ \\ 
\spec{2}{}{P}{1} (6730) & 134 & $2.24$ \\ 
\colrule 
\multicolumn{3}{c}{\spec{2}{3}{S}{1} (6897) : $ \Gamma =76.8$ keV} \T \\ 
$\spec{1}{3}{S}{1} + \pi \pi $ && $65.0$ \\ 
\spec{2}{3}{P}{0} (6692) & 201 & $7.66$ \\ 
\spec{2}{}{P}{1} (6730) & 165 & $11.5$ \\ 
$\spec{2}{}{P}{1}^\prime$ (6738) & 157 & $1.13$ \\ 
\spec{2}{3}{P}{2} (6750) & 145 & $14.6$ \\ 
\colrule 
\multicolumn{3}{c}{\spec{3}{}{D}{2} (7005) : $ \Gamma =93.7$ keV} \T \\ 
$\spec{1}{1}{S}{0} + \pi\pi$ & & $12.2$ \\
$\spec{1}{3}{S}{1} + \pi \pi $ && $9.0$ \\ 
\spec{2}{}{P}{1} (6730) & 270 & $29.1$ \\ 
$\spec{2}{}{P}{1}^\prime$ (6738) & 262 & $42.3$ \\ 
\spec{2}{3}{P}{2} (6750) & 250 & $7.24$ \\ 
\colrule 
\multicolumn{3}{c}{\spec{3}{3}{D}{1} (7006) : $ \Gamma =117$ keV} \T \\ 
$\spec{1}{3}{S}{1} + \pi \pi $ && $17.0$ \\ 
\spec{2}{3}{P}{0} (6692) & 306 & $53.4$ \\ 
\spec{2}{}{P}{1} (6730) & 270 & $25.3$ \\ 
$\spec{2}{}{P}{1}^\prime$ (6738) & 262 & $2.62$ \\ 
\spec{2}{3}{P}{2} (6750) & 251 & $1.51$ \\ 
\colrule 
\multicolumn{3}{c}{\spec{3}{3}{D}{3} (7010) : $ \Gamma =87.2$ keV} \T \\ 
$\spec{1}{3}{S}{1} + \pi \pi $ && $22.9$ \\ 
\spec{2}{3}{P}{2} (6750) & 255 & $77.0$ \\ 
\colrule 
\multicolumn{3}{c}{$\spec{3}{}{D}{2}^\prime$ (7015) : $ \Gamma =92.1$ keV} \T \\ 
$\spec{1}{1}{S}{0} + \pi\pi$ & & $9.2$ \\
$\spec{1}{3}{S}{1} + \pi \pi $ && $12.4$ \\ 
$\spec{2}{}{P}{1}^\prime$ (6738) & 272 & $37.2$ \\ 
\spec{2}{}{P}{1} (6730) & 279 & $41.0$ \\ 
\end{tabular} 
\end{ruledtabular}
\end{centering} 
\end{minipage} \hfill
\begin{minipage}{0.45\textwidth} 
\begin{centering} 
\begin{ruledtabular}
\begin{tabular}{lcc} 
Decay Mode & $k_{\gamma}$\,\text{[keV]} & Branching Fraction (\%)\B \\ 
\colrule 
\multicolumn{3}{c}{\spec{3}{3}{P}{0} (7104) : $ \Gamma =60.9$ keV}  \T\\ 
\spec{1}{3}{S}{1} (6329) & 733 & $46.4$ \\ 
\spec{2}{3}{S}{1} (6897) & 204 & $45.0$ \\ 
\spec{3}{3}{D}{1} (7006) & 97 & $8.44$ \\ 
\colrule 
\multicolumn{3}{c}{\spec{3}{}{P}{1} (7135) : $ \Gamma =87.1$ keV} \T \\ 
\spec{1}{3}{S}{1} (6329) & 761 & $32.6$ \\ 
\spec{1}{1}{S}{0} (6275) & 809 & $6.24$ \\ 
\spec{2}{3}{S}{1} (6897) & 234 & $40.4$ \\ 
\spec{2}{1}{S}{0} (6866) & 264 & $8.39$ \\ 
\spec{3}{}{D}{2} (7005) & 129 & $4.74$ \\ 
$\spec{3}{}{D}{2}^\prime$ (7015) & 119 & $4.55$ \\ 
\spec{3}{3}{D}{1} (7006) & 128 & $2.88$ \\ 
\colrule 
\multicolumn{3}{c}{$\spec{3}{}{P}{1}^\prime$ (7143) : $ \Gamma =113$ keV} \T \\ 
\spec{1}{1}{S}{0} (6275) & 816 & $32.9$ \\ 
\spec{1}{3}{S}{1} (6329) & 768 & $3.88$ \\ 
\spec{2}{1}{S}{0} (6866) & 272 & $47.0$ \\ 
\spec{2}{3}{S}{1} (6897) & 242 & $5.15$ \\ 
$\spec{3}{}{D}{2}^\prime$ (7015) & 127 & $4.22$ \\ 
\spec{3}{}{D}{2} (7005) & 137 & $6.72$ \\ 
\colrule 
\multicolumn{3}{c}{\spec{3}{3}{P}{2} (7154) : $ \Gamma =100$ keV}\T  \\ 
\spec{1}{3}{S}{1} (6329) & 777 & $35.2$ \\ 
\spec{2}{3}{S}{1} (6897) & 252 & $49.2$ \\ 
\spec{3}{}{D}{2} (7005) & 147 & $1.09$ \\ 
$\spec{3}{}{D}{2}^\prime$ (7015) & 137 & $1.18$ \\ 
\spec{3}{3}{D}{1} (7006) & 146 & $0.16$ \\ 
\spec{3}{3}{D}{3} (7010) & 142 & $13.0$ \\ 
\colrule 
\multicolumn{3}{c}{\spec{4}{}{F}{3} (7221) : $ \Gamma =77.6$ keV} \T \\ 
\spec{3}{}{D}{2} (7005) & 213 & $52.4$ \\ 
$\spec{3}{}{D}{2}^\prime$ (7015) & 203 & $42.8$ \\ 
\spec{3}{3}{D}{3} (7010) & 208 & $4.71$ \\ 
\colrule 
\multicolumn{3}{c}{\spec{4}{3}{F}{4} (7223) : $ \Gamma =79.9$ keV} \T \\ 
\spec{3}{3}{D}{3} (7010) & 210 & $100$ \\ 
\colrule 
\multicolumn{3}{c}{\spec{4}{3}{F}{2} (7233) : $ \Gamma =95.3$ keV} \T \\ 
\spec{3}{}{D}{2} (7005) & 225 & $6.73$ \\ 
$\spec{3}{}{D}{2}^\prime$ (7015) & 215 & $7.96$ \\ 
\spec{3}{3}{D}{1} (7006) & 224 & $84.8$ \\ 
\spec{3}{3}{D}{3} (7010) & 220 & $0.42$ \\ 
\colrule 
\multicolumn{3}{c}{$\spec{4}{}{F}{3}^\prime$ (7237) : $ \Gamma =89.9$ keV}  \T\\ 
$\spec{3}{}{D}{2}^\prime$ (7015) & 218 & $47.4$ \\ 
\spec{3}{}{D}{2} (7005) & 228 & $52.5$ \\ 
\end{tabular} 
\end{ruledtabular}

\vspace*{36pt}
\end{centering} 
\end{minipage} 
\label{tab:totalwidth} 
\end{table*} 
\endgroup}
\end{widetext}

\subsection{Hadronic transitions\label{subsec:pipidk}}
We evaluate the rates for hadronic transitions between $(c\bar{b})$ levels according to the prescription we detailed in \S IIIB of Ref.\,\cite{Eichten:1994gt}. The results are included in Table~\ref{tab:totalwidth}. Dipion cascades to the ground-state doublet are the dominant decay modes of \spec{2}{3}{S}{1} and \spec{2}{1}{S}{0}, and will be key to characterizing those states, as we shall discuss in \S\ref{subsec:dipicasc}.

As observed long ago by Brown and Cahn~\cite{Brown:1975dz}, an amplitude zero imposed by chiral symmetry pushes the $\pi^+\pi^-$ invariant mass distribution to higher invariant masses than phase-space alone would predict. In its simplest form, this analysis yields a universal form for the normalized dipion invariant mass distribution in quarkonium cascades $\Phi^\prime \to \Phi \,\pi^+\pi^-$,
\begin{equation}
  \frac{1}{\Gamma} \frac{d\Gamma}{d{\cal{M}}} = \text{Constant}\times
  \frac{|\vec{K}|}{M_{\Phi^\prime}^2}(2x^2-1)^2\sqrt{x^2-1} \;\;,
          \label{BandC}
\end{equation}
where $x = {\cal{M}}/2m_{\pi}$ and $\vec{K}$
is the three-momentum carried by the pion pair. The soft-pion expression
(\ref{BandC}) describes the depletion of the dipion spectrum at low invariant masses observed in the transitions
$\psi(2{S})\rightarrow \psi(1{S})\pi\pi$, $\Upsilon(2{S})\rightarrow \Upsilon(1{S})\pi\pi$, and $\Upsilon(3{S})\rightarrow
\Upsilon(2{S})\pi\pi$, but fails to account for structures in the $\Upsilon(3{S})\rightarrow \Upsilon(1{S})\pi\pi$  spectrum~\footnote{See \S7 of Ref.\,\cite{Brambilla:2004wf} and \S3.3 of Ref.\,\cite{Brambilla:2010cs} for  surveys of cascade decays.}.  We expect the $3S$
levels to lie above flavor threshold in the $(c\bar{b})$ system, and so to have very small branching fractions for cascade decays (but see the final paragraph of \S\ref{subsec:dipicasc}.

 \subsection{Properties of $(c\overline{b})$ wave functions at the origin \label{subsec:origin}}
For quarks bound in a central potential, it is convenient to separate
the Schr\"{o}dinger wave function into
radial and angular pieces, as $\Psi_{n\ell m}(\vec{r}) = R_{n\ell}(r)Y_{\ell m}(\theta,\phi)$,
where $n$ is the principal quantum number, $\ell$ and $m$ are the
orbital angular momentum and its projection, $R_{n\ell}(r)$ is the radial
wave function, and $Y_{\ell m}(\theta,\phi)$ is a spherical harmonic
\cite{[{We adopt the conventional normalization, $\int d\Omega\:Y_{\ell
m}^*(\theta,\phi)Y_{\ell^\prime m^\prime}(\theta,\phi) =
\delta_{\ell \ell^\prime}\,\delta_{m m^\prime}$.  See, e.g., the
Appendix of }][{}]BetSalt57}.  The Schr\"{o}dinger wave function is normalized, $\int{ d^3\vec{r}\, |\Psi_{n\ell m}(\vec{r})|^2} = 1$,
so that $\int_0^\infty r^2 dr |R_{n\ell}(r)| = 1$.
The value of the radial wave
function, or its first nonvanishing derivative, at the origin,
\begin{equation}
	R_{n\ell}^{(\ell)}(0)\equiv \left.\frac{d^{\,\ell}R_{n\ell}(r)}
	{dr^\ell}\right|_{r=0}\;\;\;,
	\label{wvfc}
\end{equation}
is required to evaluate pseudoscalar decay constants and production rates
through heavy-quark fragmentation.  Our calculated values of $|R_{n\ell}^{(\ell)}(0)|^2$ are given in Table~\ref{tab:psizero}.
 {\begingroup
\begin{table}[bth]
\caption{Squares of radial wave functions at the origin and related quantities (cf. \eq{wvfc}) for
$(c\bar{b})$ mesons. \label{tab:psizero}}
\begin{tabular}{cc}
\toprule
Level & $|R_{n\ell}^{(\ell)}(0)|^2$\T\B\\
\colrule
$1S$ & $1.994\gev^{3}$\T \\[1pt]
$2P$ &  $0.3083\gev^{5}$\\
$2S$ &  $1.144\gev^{3}$\\[1pt]
$3D$ &  $0.0986\gev^{7}$\\
$3P$ &  $0.3939\gev^{5}$\\
$3S$ & $0.9440\gev^{3}$\\[1pt]
$4F$ & $0.0493\gev^{9}$\\
$4D$ & $0.1989\gev^{7}$\\
$4P$ & $0.4540\gev^{5}$ \\
$4S$ & $0.8504\gev^{3}$\\
\botrule
\end{tabular}
\end{table}
\endgroup}

The pseudoscalar decay constant
$f_{B_c}$, which enters the calculations of annihilation
decays such as $c\bar{b}\rightarrow W^+ \rightarrow \tau^+ + \nu_\tau$, is defined by
\begin{equation}
\langle 0|A_\mu(0)|B_c(q)\rangle = i f_{B_c} \mathsf{V}_{cb} \,q_\mu\;\;\; ,
\end{equation}
where $A_\mu$ is the axial-vector part of the charged weak current,
$\mathsf{V}_{cb}$ is an element of the Cabibbo-Kobayashi-Maskawa quark-mixing
matrix, and $q_\mu$ is the four-momentum of the $B_c$.  
Its counterpart for the vector state is
\begin{equation} 
\langle 0|V_\mu(0)|B^*_c(q)\rangle = i f_{B^*_c} \mathsf{V}_{cb} \varepsilon^*_\mu\;\;\; ,
\label{eq:Vecdkdef}
\end{equation}
where $V_\mu$ is the vector part of the charged weak current and $\varepsilon^*_\mu$ is the polarization vector of the $B^*_c$. 
The ground-state pseudoscalar and vector decay constants are given in terms of the wave function at the origin by the Van Royen--Weisskopf formula~\cite{[{}][{ The factor of 3 accounts for quark color.}]VanRoyen:1967nq},
generically
\begin{equation}
f^2_{B_c^{(*)}} = \frac{3|R_{10}(0)|^2}{\pi M}\overline{C}^2(\alphas),
\label{eq:decayconst}
\end{equation}
where the leading-order QCD correction is given by~\cite{Braaten:1995ej, *Berezhnoy:1996an}
\begin{equation}
\overline{C}^2(\alphas)= 1- \frac{\alphas}{\pi}\left(\delta^{\text{P,V}} - \frac{m_c - m_b}{m_c+m_b}\ln{\frac{m_c}{m_b}}\right),
\label{eq:vRWQCD}
\end{equation}
and
\begin{equation}
\delta^{\text{P}} = 2; \quad \delta^{\text{V}} = 8/3.
\label{eq:deltadef}
\end{equation}
Choosing the representative value $\alphas = 0.38$, and using the quark masses given in \eq{eq:Cornellparams}, we find 
\begin{equation}
\overline{C}(\alphas) = 
\begin{array}{c}
0.904, \text{ P} \\[3pt]
0.858, \text{ V}
\end{array}
\label{eq:radcorPV}
\end{equation}
Consequently, we estimate the ground-state meson decay constants as
\begin{equation}
f_{B_c} = 498\mev; \quad f_{B_c^*} = 471\mev,
\label{eq:fvals}
\end{equation}
so that $f_{B_c^*}/f_{B_c} = 0.945$.  The compact size of
the $(c\bar{b})$ system enhances the pseudoscalar decay constant relative to $f_\pi$ and $f_K$.

This is to be compared to a state-or-the-art lattice evaluation~\cite{[{}][{. For further work on semileptonic decays, see }]Colquhoun:2015oha,*Lytle:2016ixw,*Lytlemicro},
$f_{B_c} = 434 \pm 15\mev$,
which entails improved NonRelativistic QCD for the valence $b$ quark and the Highly Improved Staggered Quark (HISQ) action for the lighter quarks on gluon field configurations that include the effect of $u/d$, $s$ and $c$ quarks in the sea with the $u/d$ quark masses going down to physical values. The same calculation yields $f_{B_c^*}/f_{B_c} = 0.988 \pm 0.027$. A calculation in the framework of QCD sum rules gives $f_{B_c} = 528 \pm 19\mev$~\cite{Baker:2013mwa}. 

\section{Production of (${c\bar{b}}$) states at the Large Hadron Collider\label{sec:lhcprod}}
We present in Table~\ref{tab:prod} cross sections for the production of $B_c$ states at the Large Hadron Collider, calculated using the framework of the  BCVEGPY2.2 generator~\cite{[{}][{. We use (derivatives of) wave functions at the origin derived from our current work. The quark mass parameters in this program vary with the produced state, to reproduce its mass. $1S\text{: }m_b = 5.000, m_c = 1.275;\; 2S\text{: }m_b = 5.234,  m_c = 1.633;\; 2P\text{: }m_b = 5.184,  m_c = 1.573;\; 3S\text{: }m_b = 5.447,  m_c = 1.825;\; 3P\text{: }m_b = 5.502, m_c =  1.633$, all in GeV. }]Chang:2005hq}, which we have extended to include the production of $3P$ states. Cross sections for the physical $\spec{(2,3)}{}{P}{1}^{(\prime)}$ states are appropriately weighted mixtures of the \spec{}{3}{P}{1}  and \spec{}{1}{P}{1} cross sections.
 {\begingroup
\begin{table}[tbh]
\caption{Production rates (in  nb) for $(c\bar{b})$ states in $pp$ collisions at the LHC. 
The production rates were calculated using the BCVEGPY2.2 generator of Ref.\,\cite{Chang:2005hq}, extended to include the production of $3P$ states.
 Color-octet contributions to $s$-wave production are small; we show them (following $\mid$) only for the $1S$ states.\label{tab:prod} }
 \begin{tabular}{cccc}
 \toprule
$(c\bar{b})$ level & $\sigma(\sqrt{s}=8\tev)$ &  $\sigma(\sqrt{s}=13\tev)$ & $\sigma(\sqrt{s}=14\tev)$\T\\[2pt]
\colrule 
\spec{1}{1}{S}{0} & $ 46.8 \mid 1.01 $  &  $80.3 \mid 1.75$ & $88.0 \mid 1.90$\T \\    
\spec{1}{3}{S}{1} &   $123.0 \mid 4.08$  &  $219.1 \mid 6.97$ & $237.0 \mid 7.55$ \\ 
\spec{2}{3}{P}{0} &  $1.113$  &  $1.959 $ & $2.108$ \\ 
\spec{2}{3}{P}{1} & $2.676$  &  $4.783$  & $5.214$  \\ 
\spec{2}{1}{P}{1} &  $3.185$  &  $5.702$  & $6.166$\\ 
\spec{2}{3}{P}{2} & $6.570$   &  $11.57$  &  $12.64$ \\ 
\spec{2}{1}{S}{0} & $9.58$  &  $16.94$  & $18.45$ \\ 
\spec{2}{3}{S}{1} & $23.46$   &  $41.72$ & $45.53$ \\ 
\spec{3}{3}{P}{0} & $0.915$    &  $1.642$  & $1.806$ \\ 
\spec{3}{3}{P}{1} & $2.263$    &  $4.082$  & $4.478$ \\ 
\spec{3}{1}{P}{1} &  $2.695$    &  $4.817$  & $5.287$ \\ 
\spec{3}{3}{P}{2} &  $5.53$    &  $9.98$  & $10.90$ \\ 
\spec{3}{1}{S}{0} & $4.23$  &  $7.53$  & $8.08$ \\ 
\spec{3}{3}{S}{1} & $10.16$   &  $18.21$  & $19.83$ \\[1pt]
\botrule
\end{tabular}
\end{table}
\endgroup}

\begin{figure}[tb] 
      \includegraphics[width=\columnwidth]{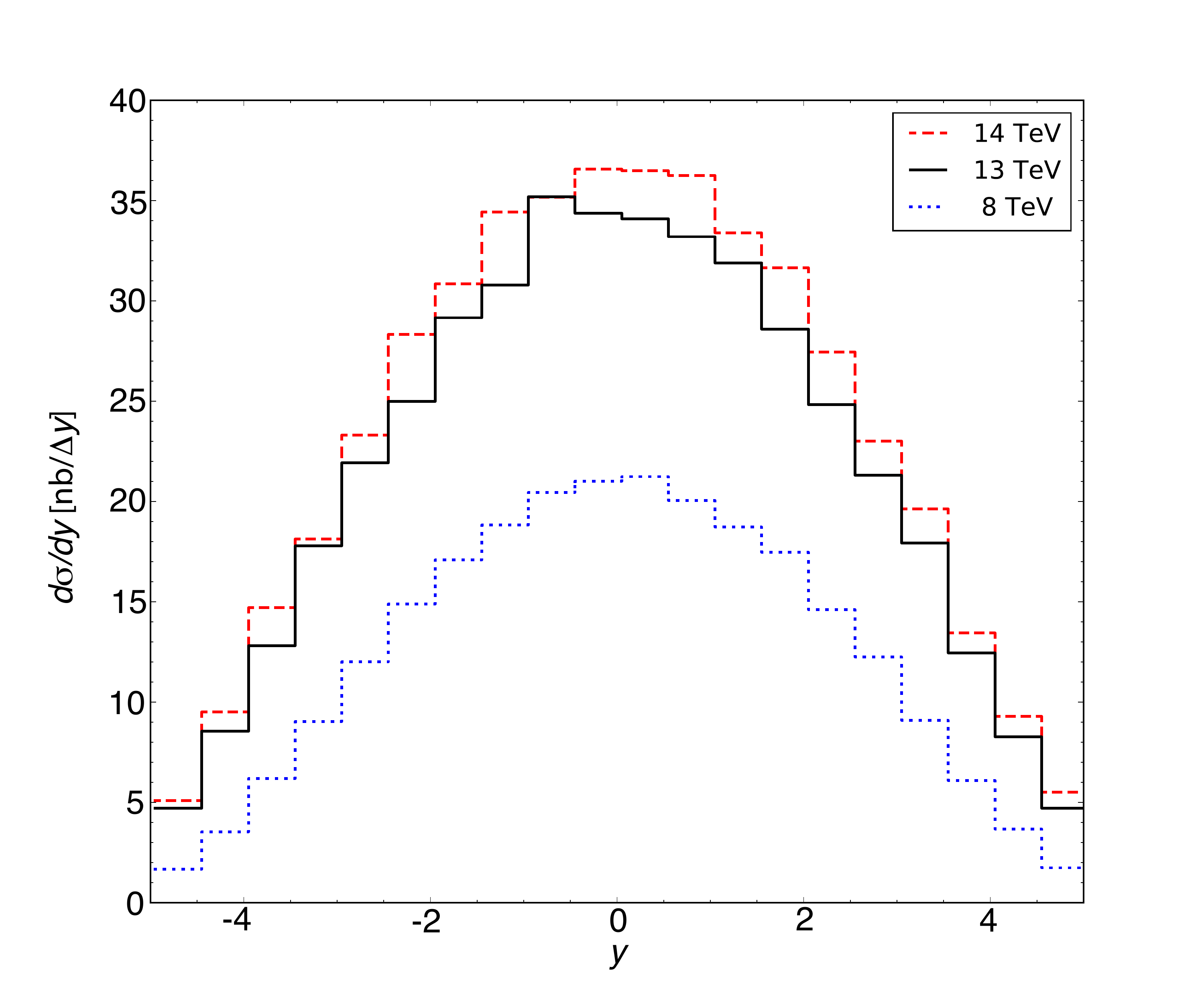} 
      \caption{Rapidity distribution for  production of  $B_c^*$  in $pp$ collisions at  $\sqrt{s} =8\tev$ (dotted blue curve), $\sqrt{s} =13\tev$ (solid black curve), and $\sqrt{s} = 14\tev$ (dashed red curve), calculated using BCVEGPY2.2~\cite{Chang:2005hq}. The bin width is $\Delta y = 0.5$. The mild asymmetries are statistical fluctuations.
      \label{fig:yenergy}}
    \end{figure}

\begin{figure}[tb] 
   \centering
   \includegraphics[width=\columnwidth]{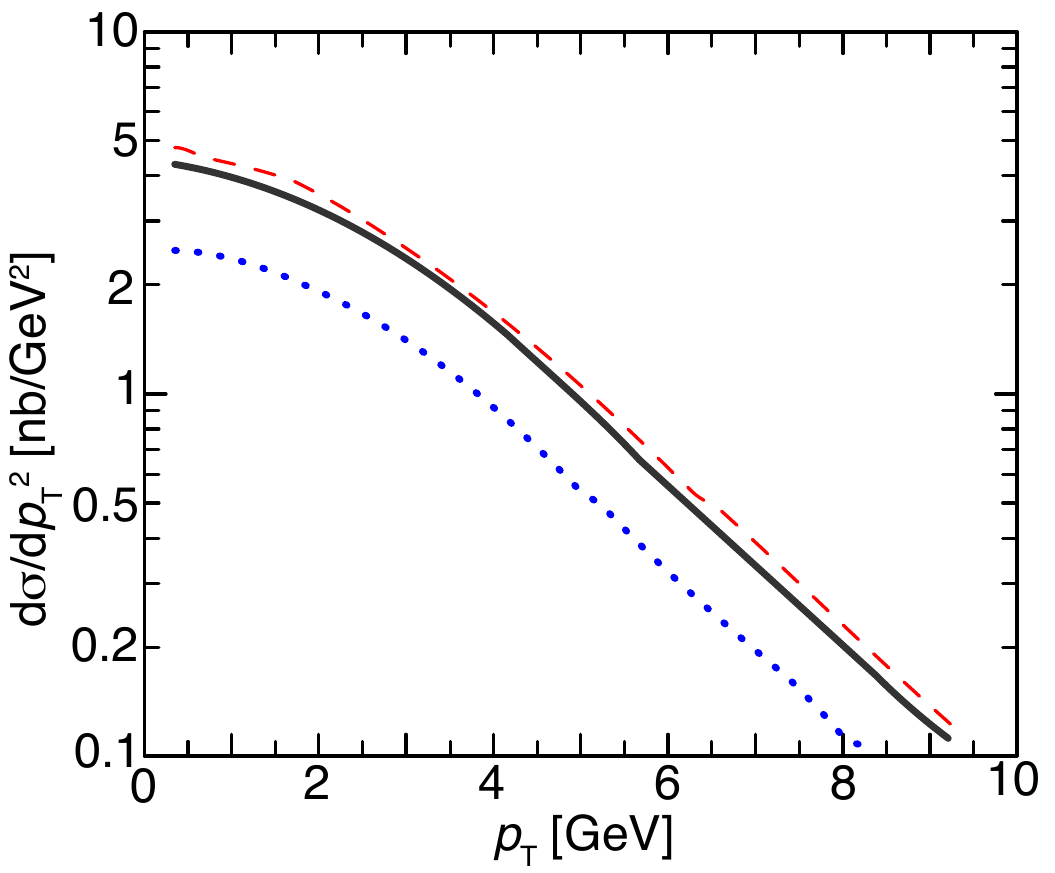} 
   \caption{Transverse momentum distribution of  $B_c$ produced in $pp$ collisions at  $\sqrt{s} =8\tev$ (dotted blue curve), $\sqrt{s} =13\tev$ (solid black curve), and $\sqrt{s} = 14\tev$ (dashed red curve), calculated using BCVEGPY2.2~\cite{Chang:2005hq} Small shape variations are statistical fluctuations.
   \label{fig:ptdist}}
\end{figure}

\begin{figure}[tb] 
       \includegraphics[width=\columnwidth]{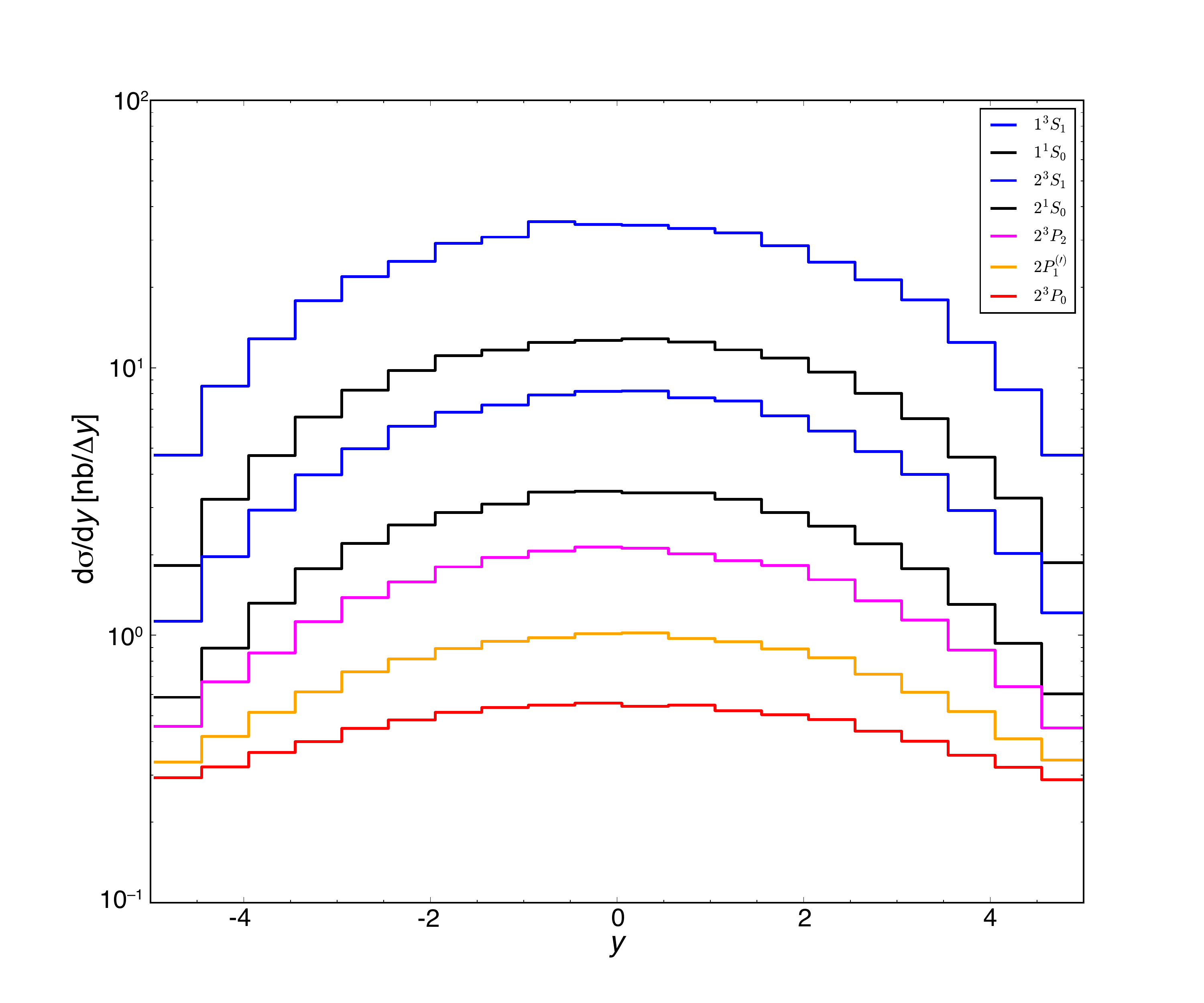} 
       \caption{Rapidity distributions for the production of  low-lying $(c\bar b)$ states in $pp$ collisions at  $\sqrt{s} =13\tev$, calculated using BCVEGPY2.2~\cite{Chang:2005hq}.  From highest to lowest, the histograms refer to production of the   \spec{1}{3}{S}{1}, \spec{1}{1}{S}{0}, \spec{2}{3}{S}{1}, \spec{2}{1}{S}{0}, \spec{2}{3}{P}{2}, \spec{2}{}{P}{1^{(\prime)}}, \spec{2}{3}{P}{0} levels. 
       \label{fig:ystates}}
    \end{figure}

The rapidity distributions (for $B_c^*$ production, Figure~\ref{fig:yenergy}) and transverse-momentum distributions (shown for $B_c$ production, Figure~\ref{fig:ptdist})  are similar in character for  $\sqrt{s} = 8, 13, \text{ and } 14 \tev$. 
 The rapidity distributions for low-lying $(c\bar{b})$ states are shown in Figure \ref{fig:ystates}.  The acceptance of the CMS and ATLAS detectors covers central pseudorapidity $|\eta| \le 2.5$, whereas the geometrical acceptance of the LHC$b$ detector is characterized by $2 \le \eta \le 5$.
For comparison, approximately $68\%$ of the $B_c^*$ cross section lies within $|y| \le 2.5$, and approximately $22\%$ is produced at forward rapidities $y > 2$. Similar fractions hold for all the $(c\bar{b})$ levels.

\subsection{Dipion cascades\label{subsec:dipicasc}}

The path to establishing excited states will proceed by resolving two separate peaks in the invariant mass distributions associated with the cascades $B_c^{\prime}\rightarrow B_c \pi^+ \pi^-$ and $B_c^{*\prime} \rightarrow B_c^* + \pi^+ \pi^- $, $B_c^* \rightarrow B_c +\slashi{\gamma}$ (gamma unobserved). The splitting between the peaks is set by the difference of mass differences, 
\begin{equation}
\Delta_{21}\equiv[M(B_c^{*\prime}) - M(B_c^{\prime})] - [M(B_c^{*}) - M(B_c)], 
\label{eq:Del21def}
\end{equation}
generically expected to be negative~\cite{[{In an effective power-law potential $V(r) = \lambda r^\nu$, $\Delta_{21} <0$ so long as $\nu <1$. See \S4.1.1 and \S5.3.2 of }][{ particularly Eqns.\ (4.21, 4.22).}]Quigg:1979vr}. The corresponding quantity is approximately $-64\mev$ in the $(c\bar{c})$ family and $-37\mev$ in the $(b\bar{b})$ family~\cite{Tanabashi:2018oca}. 
For the $(c\bar{b})$ system, a modern lattice simulation~\cite{Dowdall:2012ab} gives $\Delta_{21} = -15\mev$, whereas the result of our potential-model calculation is $-23\mev$.  In these circumstances, the undetected four-momentum of the photon means that the reconstructed ``$B_c^{*}$'' mass should correspond to the lower peak. 

\begin{figure}[tb] 
   \centering
   \includegraphics[width=\columnwidth]{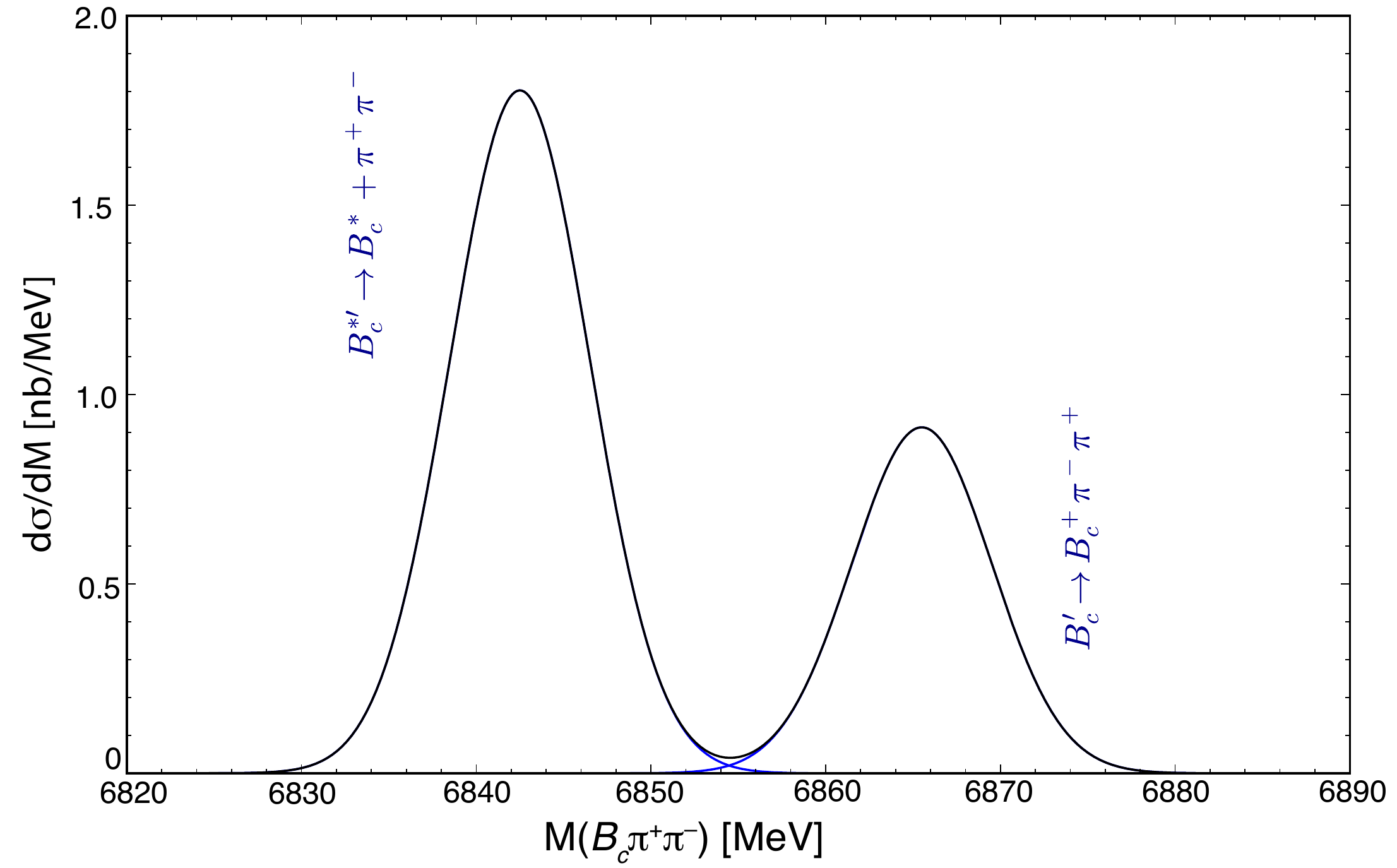} 
   \caption[]{Calculated positions and relative strengths of the two-pion cascades $B_c(2S) \to B_c(1S)\pi^+\pi^-$, represented as Gaussian line shapes with standard deviation of $4\mev$. Production rates are given in Table~\ref{tab:prod} and  branching fractions  in  Table \ref{tab:totalwidth}. We assume that the photon in the transition $B_c^* \rightarrow B_c  + \gammiss$ is not included in the reconstruction. Rates confined to rapidity $|y| \le 2.5$ are $0.68 \times$ those shown.  \label{fig:pitrans}}
\end{figure}

We show an example of what is to be expected in Figure~\ref{fig:pitrans}, 
taking the direct production cross sections (with no rapidity cuts) from Table~\ref{tab:prod} and the branching fractions from Table~\ref{tab:totalwidth}.  The (relative heights of, relative number of events in) the peaks measures the ratio 
\begin{equation}
\mathcal{R} \equiv \frac{\sigma(B_c^{*\prime}+X)~\BR(B_c^{*\prime} \rightarrow B_c^* + \pi^+ \pi^-)}{\sigma(B_c^{\prime}+X)~\BR(B_c^{\prime} \rightarrow B_c + \pi^+ \pi^-)}.
\label{eq:peakheights}
\end{equation}
At $\sqrt{s} = 13\tev$, the ratio of cross sections is nearly 2.5. Taking account of the branching fractions, we estimate $\mathcal{R} \approx 2$.  If $B_c^*$ and $B_c$ were produced with equal frequency, we would find $\mathcal{R} \approx 0.8$.

Now the CMS Collaboration~\cite{Sirunyan:2019osb} at the Large Hadron Collider, analyzing $140\fb^{-1}$ of $pp$ collisions at $\sqrt{s} = 13\tev$, has observed a pattern that closely resembles the template of Figure~\ref{fig:pitrans}. In the distribution of $M(B_c\pi^+\pi^-) - M(B_c)^{\text{obs}}+M(B_c)$, they reconstruct a peak at $6871.0 \pm 1.2\mev$, which they identify as $B_c(2S)$, and a second peak  $29.0 \pm 1.5\mev$ lower in mass (statistical errors only). [The observed $B_c$ mass is replaced, event by event, with the world-average value to sharpen resolution.] The putative $B_c(2S)$ lies within $5\mev$ of our expectation for the \spec{2}{1}{S}{0} level, and the separation is to be compared with our expectation of $23\mev$. If we impose the scaling relation \eq{eq:hfsrat} for the hyperfine splittings, we reproduce the observed 29-MeV separation with $M(B_c^*) - M(B_c) = 68\mev$, $M(B_c^{*\prime}) - M(B_c^\prime) = 39\mev$. The $B_c^* \to B_c + \gamma$ photon momentum would be $68\mev$.

An unbinned extended maximum-likelihood fit to the CMS data returns $66 \pm 10$ events for the lower peak and $51 \pm 10$ for the upper. These yields are not yet corrected for detection efficiencies and acceptances, so they cannot be used to infer ratios of production cross sections times branching fractions. We look forward to the final result and to studies of the $\pi^+\pi^-$ invariant mass distribution as next steps in $B_c$ spectroscopy.

Our calculations indicate that the \spec{3}{}{S}{} levels will lie above flavor threshold (see \S\ref{subsec:openflav}, especially the discussion surrounding Figures\,\ref{fig:decay1s0} and \ref{fig:decay3s1}), but it is conceivable that coupled-channel effects might push one or both states lower in mass. For that reason, it is worth examining the $B_c\pi^+\pi^-$ mass spectrum up through $7200\mev$ for indications of $\spec{3}{1}{S}{0} \to B_c\pi^+\pi^-$ and $\spec{3}{3}{S}{1} \to B_c^*\pi^+\pi^-$ lines. According to our estimate of the \spec{3}{}{S}{} hyperfine splitting, the \spec{3}{3}{S}{1} line would lie about $28\mev$ below the \spec{3}{1}{S}{0} line ($36\mev$ if we reset the \spec{1}{}{S}{} splitting to $68\mev$). For orientation, note that $\mathcal{B}(\Upsilon(3S) \to \Upsilon(1S)\pi^+\pi^-) = 4.37 \pm 0.08\%$, while $36\%$ of $\Upsilon(3S)$ decays proceed through the $ggg$ channel, which is not available to the $(c\bar{b})$ states. According to Table\,\ref{tab:prod}, the \spec{3}{}{S}{} states are produced at approximately $44\%$ of the rate for their \spec{2}{}{S}{} counterparts.

\subsection{Electromagnetic transitions\label{subsec:e1casc}}

It may in time become possible for experiments to detect some of the more energetic E1-transition photons that appear in Table~\ref{tab:totalwidth}. As an incentive for the search, we show in Figure~\ref{fig:e1trans}
\begin{figure}[tb] 
   \includegraphics[width=\columnwidth]{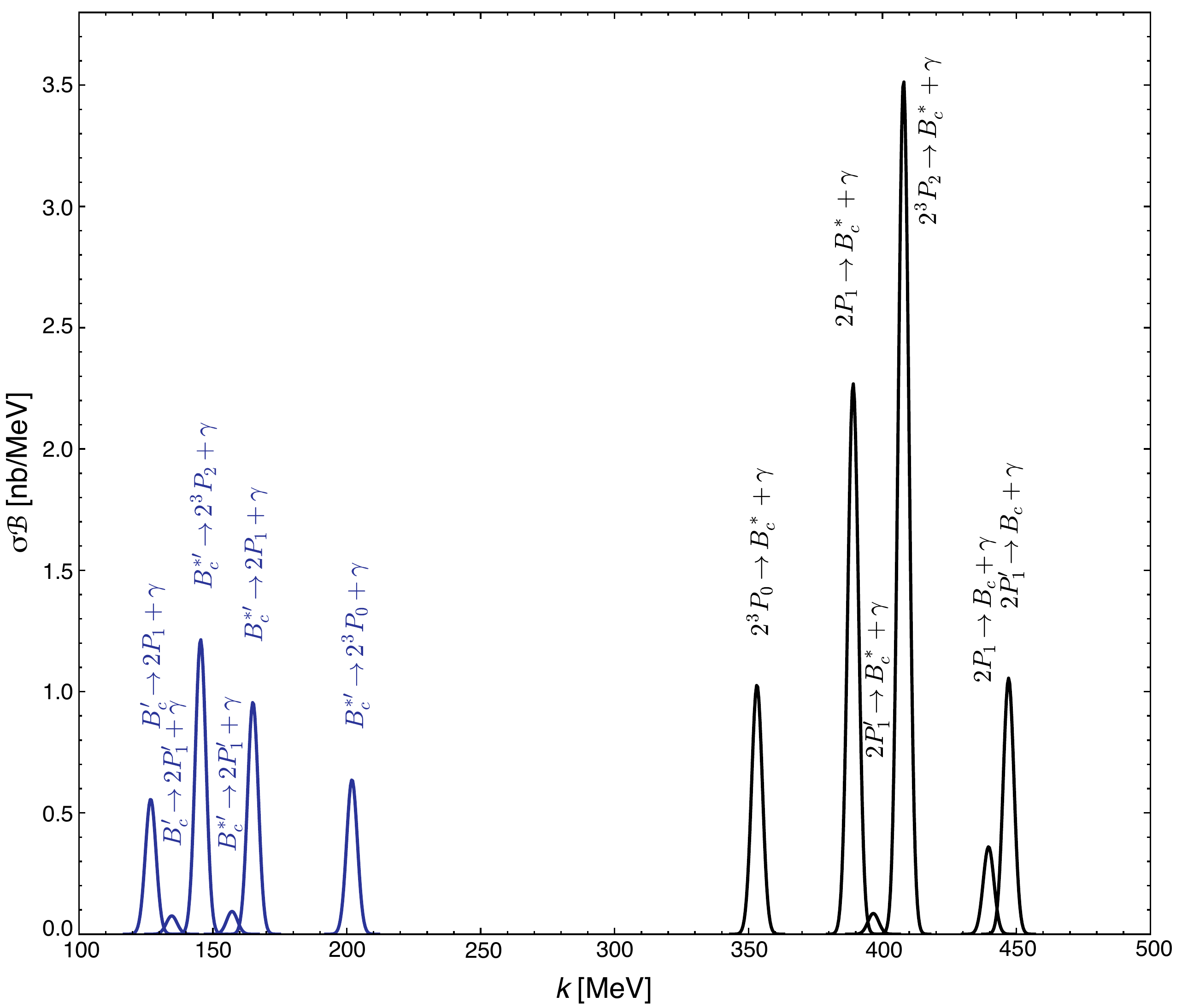} 
   \caption[]{Photon energies $k$ and relative strengths of  E1 transitions from $2S \to 2P$ (left group, blue curves) and $2P \to 1S$ (right group, black curves) $(c\bar b$) states. Production rates are taken from Table~\ref{tab:prod} and  branching fractions from 
   Table~\ref{tab:totalwidth}. We suppose that the photon transition $B_c^* \rightarrow B_c  + \gammiss$ goes unobserved in the cascade transitions. We assume Gaussian lineshapes with standard deviation $2\mev$.  \label{fig:e1trans}}
\end{figure}
the spectrum of E1 photons in decays of the \spec{2}{3}{S}{1}\ and \spec{2}{1}{S}{0}\ levels as well as the $2P \to 2S$ transitions, assuming as always a missing $B_c^* \to B_c \gammiss$ photon in the reconstruction. Here we include direct production of the $2P$ states as well as feed-down from $2S \to 2P$ transitions. The strong $B_c^* \to B_c$ line arising from direct production of $B_c^*$, for which we calculate $\sigma\cdot\mathcal{B} \approx 225\nb$ at $\sqrt{s} = 13\tev$, is probably too low in energy to be observed. More promising are the $2P$ levels, which might show themselves in $B_c + \gamma$ invariant mass distributions. These lines make up the right-hand group (black lines) in Figure~\ref{fig:e1trans}. The $\spec{2}{3}{P}{2}(6750) \to B_c^*\gamma$ line is a particularly attractive target for experiment, because of the favorable production cross section, branching fraction, and 409-MeV photon energy. The  $2P$ masses inferred from transitions to $B_c^*$ will be shifted downward because of the unobserved M1 photon. It is not possible to produce enriched samples of the $2S$ levels by tuning the energy of $e^+e^-$ collisions, as is done for $\jpsi$ and $\Upsilon$, so reconstruction of the left-hand group of $2S \to 2P$ transitions (blue lines in Figure~\ref{fig:e1trans}) will be problematic.

In the far future, combining  the photon transition energies and relative rates with expectations for production and decay may eventually  make it possible to disentangle mixing of the spin-singlet and spin-triplet $J = L$ states.

\subsection{States above open-flavor threshold\label{subsec:openflav}}
We estimate the strong decay rates for $(c\bar{b})$ states that lie above flavor threshold using the 
Cornell coupled-channel formalism~\cite{Eichten:1978tg, *Eichten:1979ms} that we elaborated and applied to charmonium states in~\cite{Eichten:2004uh,Eichten:2005ga}.

\begin{figure}[btp] 
   \includegraphics[width=\columnwidth]{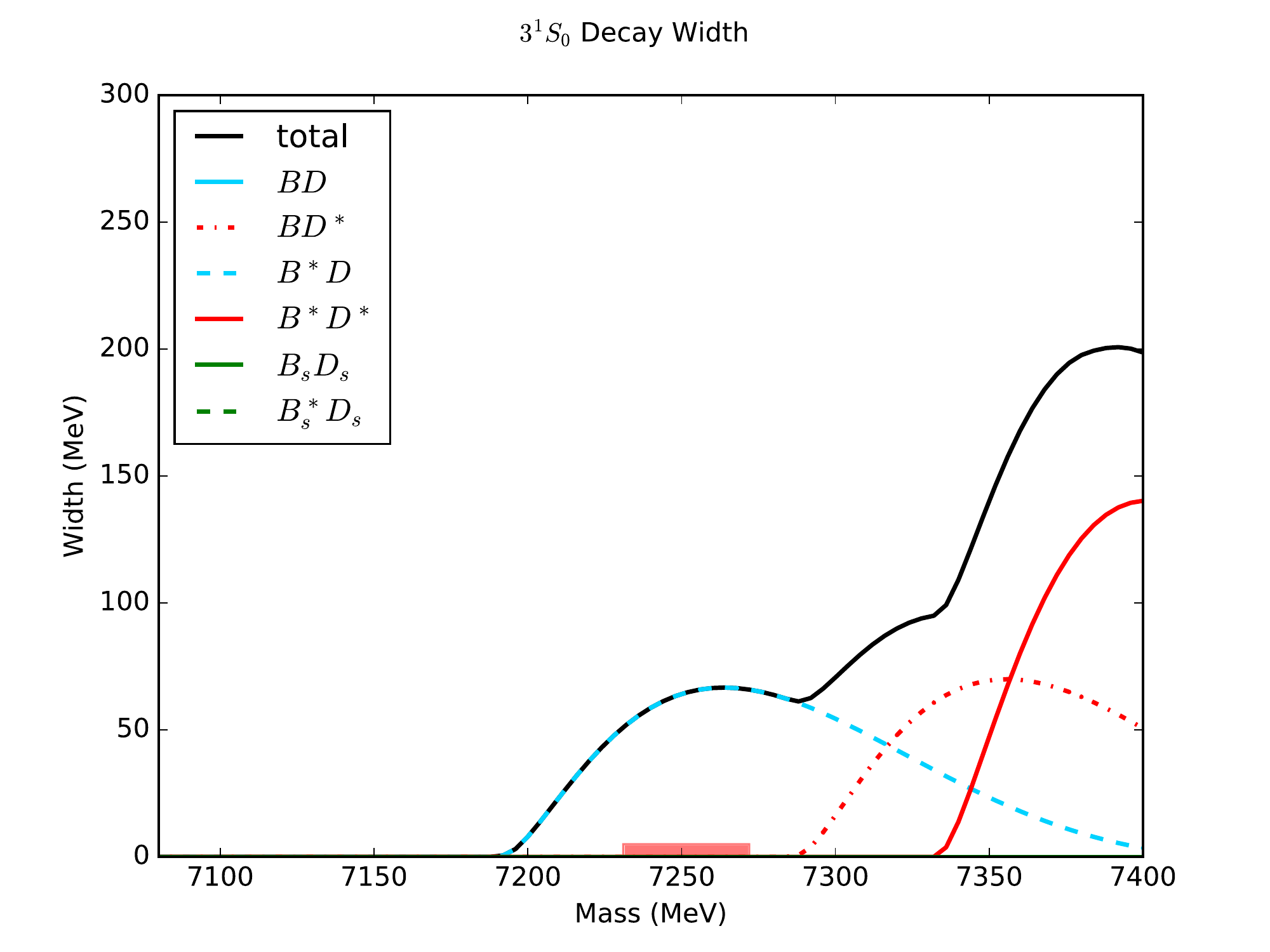} 
      \caption{Strong decay widths of the \spec{3}{1}{S}{0} $(c\bar{b})$ level near open-flavor threshold.  The shaded band on the mass axis indicates $\pm 20\mev$ around our nominal value for the mass of this state, $7253\mev$. \label{fig:decay1s0}}
\end{figure}

\begin{figure}
   \includegraphics[width=\columnwidth]{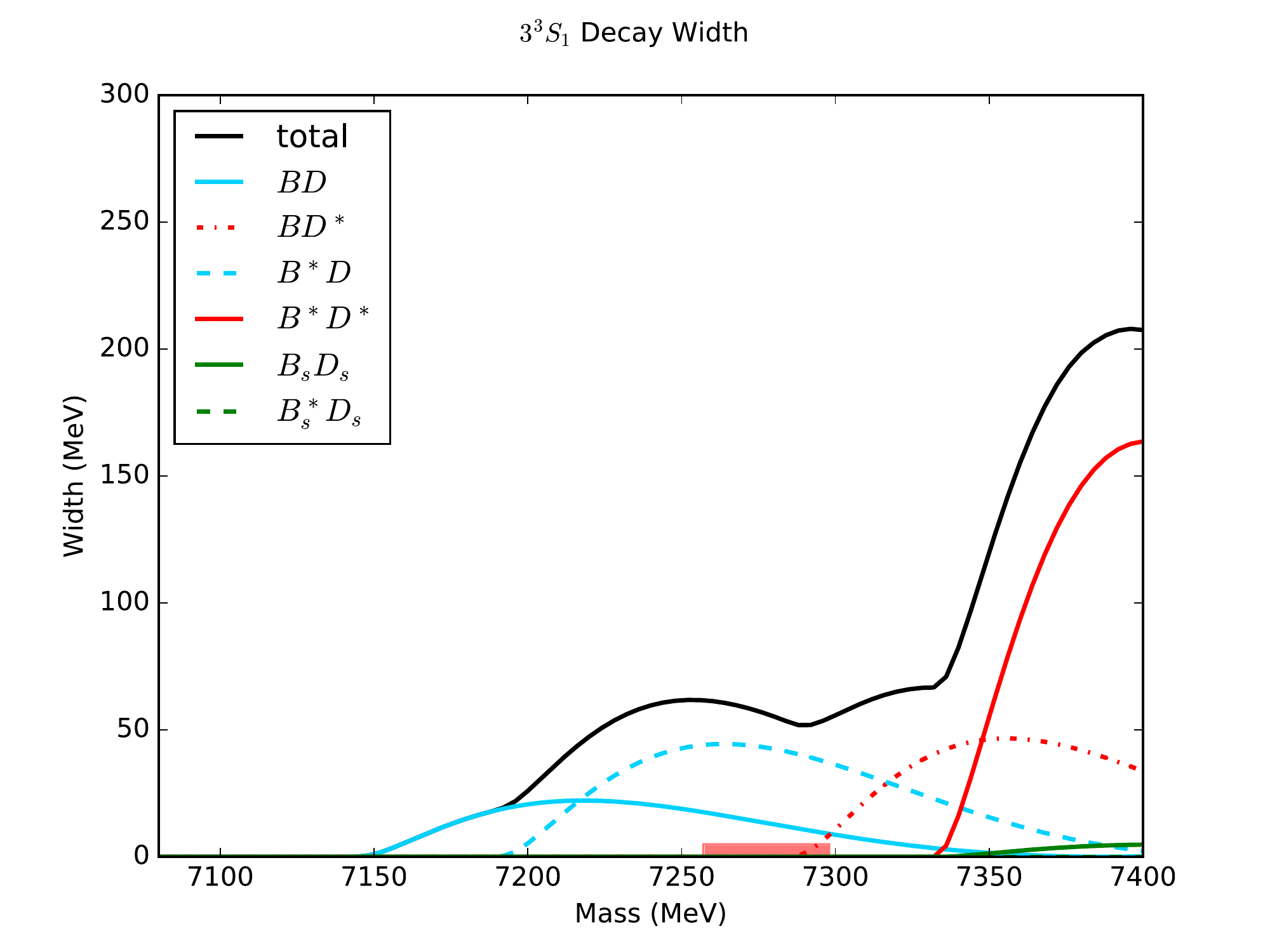} 
 \caption{Strong decay widths of the \spec{3}{3}{S}{1} $(c\bar{b})$ level near open-flavor threshold.  The shaded band on the mass axis indicates $\pm 20\mev$ around our nominal value for the mass of this state, $7279\mev$.\label{fig:decay3s1}}
  \end{figure}

 We expect both the  \spec{3}{1}{S}{0} and \spec{3}{3}{S}{1} states to lie above threshold for strong decays.  The \spec{3}{1}{S}{0} state can decay into the final state $B^*\!D$  and the  \spec{3}{3}{S}{1} level has decays into both the $BD$ and 
$B^*\!D$ final states. The open decay channels as a function of the masses of these states is shown in Figures~\ref{fig:decay1s0} and \ref{fig:decay3s1}.

\begin{figure}[tbp] 
   \centering
   \includegraphics[width=\columnwidth]{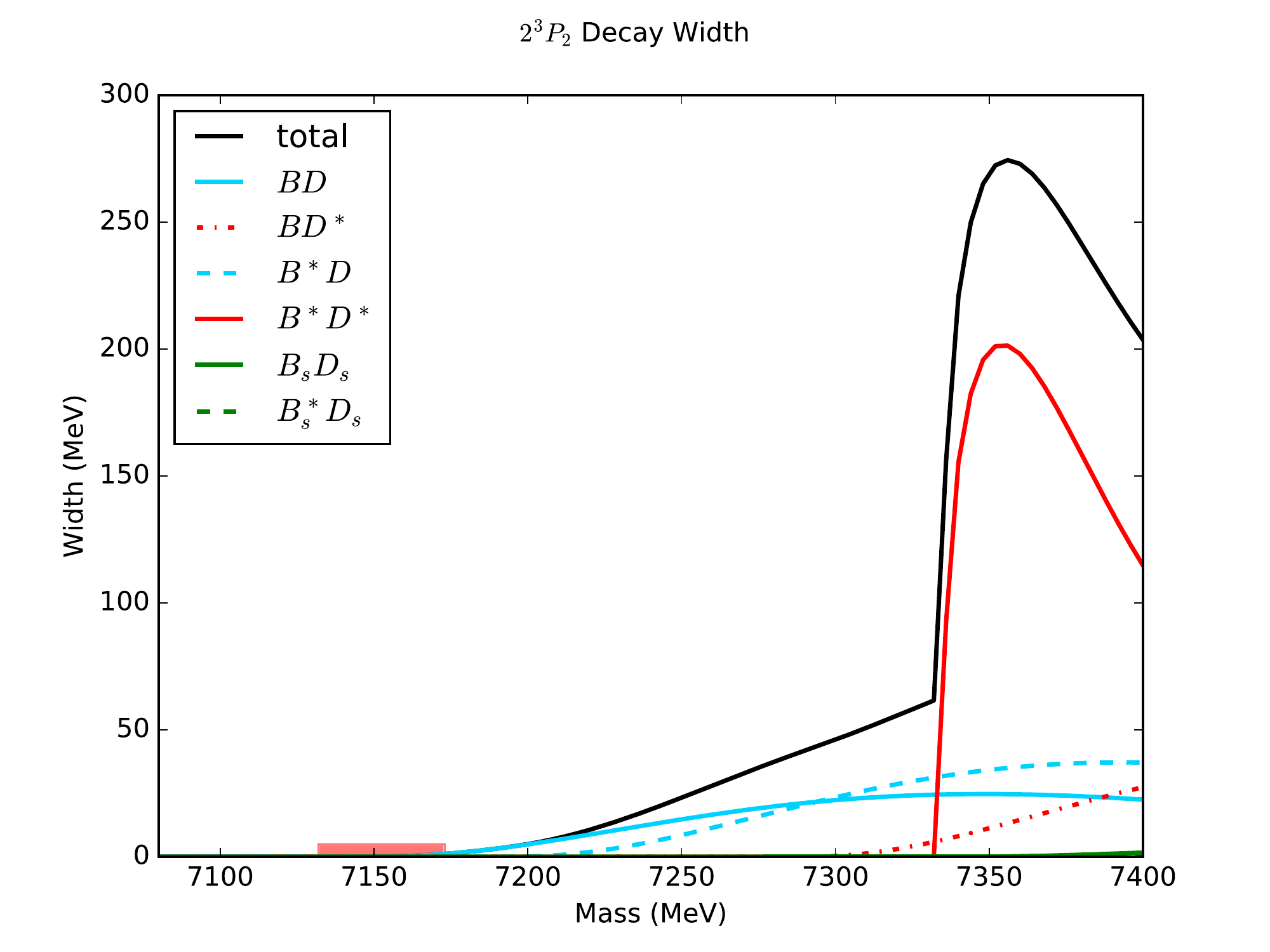} 
   \caption{Strong decay widths of the \spec{3}{3}{P}{2} $(c\bar{b})$ level near open-flavor threshold. The shaded band on the mass axis indicates $\pm 20\mev$ around our nominal value for the mass of this state, $7154\mev$.   \label{fig:decay3p2}}
\end{figure}

The \spec{3}{3}{P}{2} state might be observed as a very narrow ($d$-wave) $BD$ line near open-flavor threshold.  Its decay width  as a function of mass for the $2P$ states are given in
Figure~\ref{fig:decay3p2}. 

\begin{figure}[tb] 
    \includegraphics[width=\columnwidth]{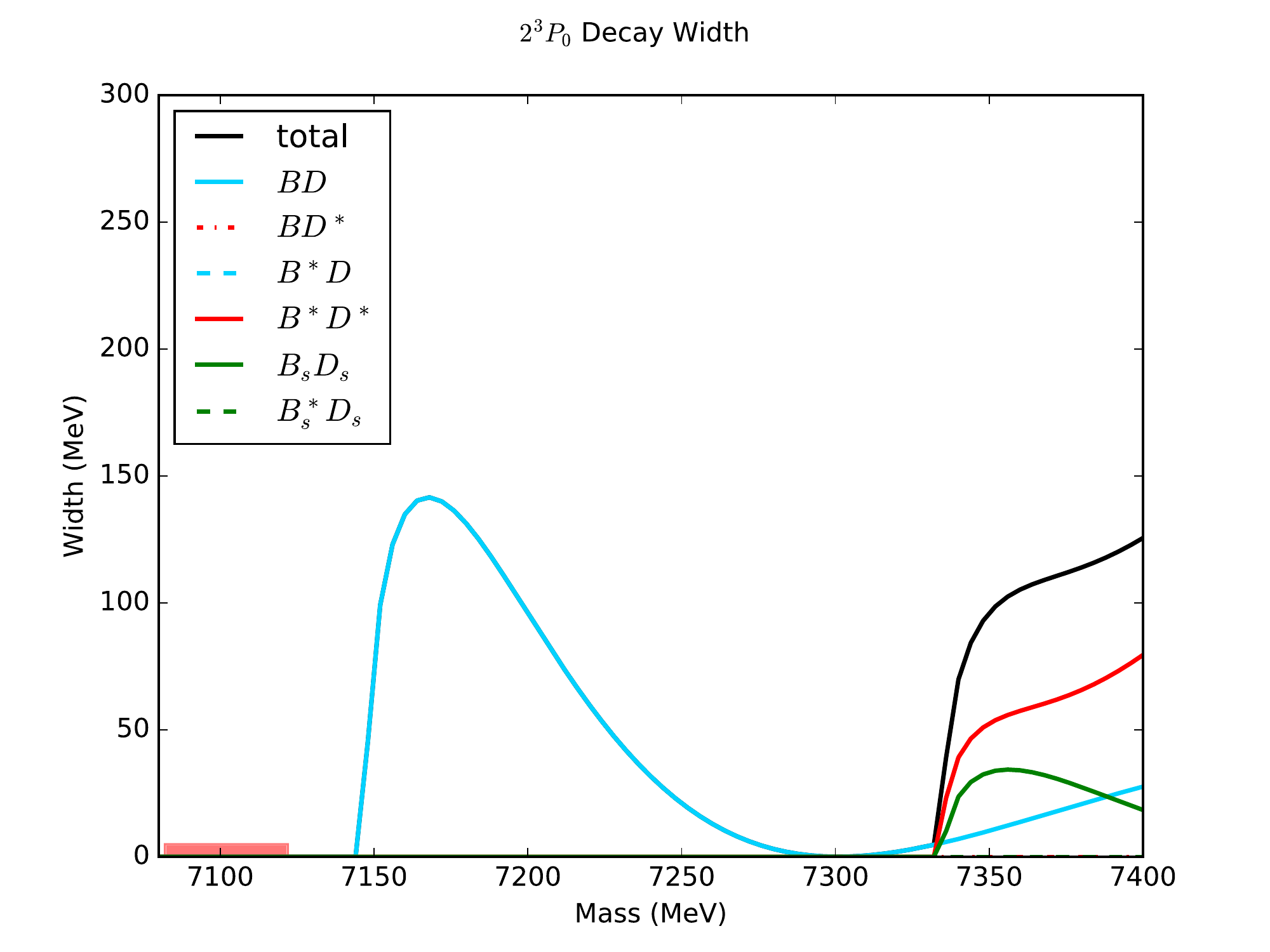} 
   \caption{Estimated strong decay widths of the \spec{3}{3}{P}{0} $(c\bar{b})$ level near open-flavor threshold.  The shaded band on the mass axis indicates $\pm 20\mev$ around our nominal value for the mass of this state, $7104\mev$. \label{fig:decay3p0}}
\end{figure}

\begin{figure}[tbh]
    \includegraphics[width=\columnwidth]{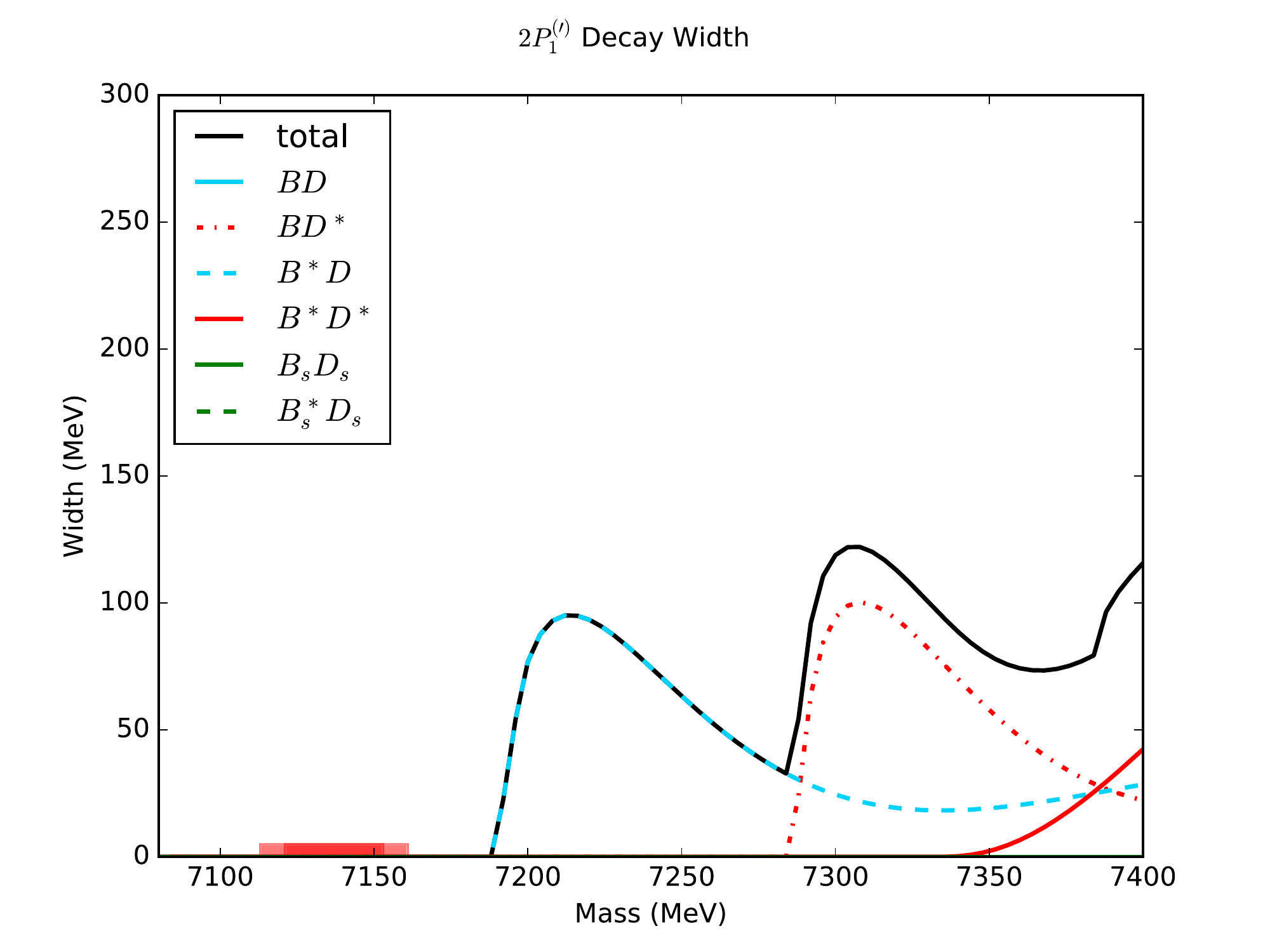} 
   \caption{Strong decay widths of the $3P_1$ or $3P_1^{(\prime)}$.  The shaded band on the mass axis indicates $\pm 20\mev$ around our nominal values for the masses of these state, $7135\text{ and }7143\mev$. \label{fig:decayp1}}
\end{figure}

In the phenomenological models the remaining $3P$ states lie just below the thresholds for strong
decays.  However they are near enough to these thresholds that there might be interesting behavior at the threshold for $B^*D$ in the $\spec{3}{}{P}{1}^{(\prime)}$ cases and for the  $BD$ 
threshold in the case of the \spec{3}{3}{P}{0} state.  Figure\,\ref{fig:decay3p0} shows that the  \spec{3}{3}{P}{0} width grows  rapidly just above threshold.  The strong decay widths as a function of mass for the $3P_1$ and $3P_1^\prime$ states have a common behavior, displayed  in Figure~\ref{fig:decayp1}. 

It is worth keeping in mind that while narrow $BD$ peaks may signal excited $(c\bar{b})$ levels, narrow $\bar{B}D$ peaks could indicate nearly bound $bc\bar{q}_k\bar{q}_l$ tetraquark states~\cite{[{}][{ and references cited therein.}]Eichten:2017ffp}.


\section{Tera-$Z$ Prospects \label{sec:teraz}}
In response to the discovery of the 125-GeV Higgs boson, $H(125)$~\cite{Aad:2012tfa,*Chatrchyan:2012xdj}, plans for large circular electron--positron colliders  (FCC-ee~\cite{tlep} and CEPC~\cite{cepc}) are being developed as $e^+e^- \to HZ^0$ ``Higgs factories'' to run at c.m.\ energy $\sqrt{s} \approx 240\gev$. As now envisioned, these machines would have the added capability of high-luminosity running at $\sqrt{s} = M_Z$ that would accumulate $10^{12}$ examples of the reaction $e^+e^- \to Z^0$. With the observed branching fraction, $\mathcal{B}(Z^0 \to b\bar{b}) = (15.12 \pm 0.05)\%$~\cite{Tanabashi:2018oca}, the tera-$Z$ mode would produce some $3 \times 10^{11}$ boosted $b$-quarks, which would enable high-sensitivity searches for $(c\bar{b})$ states in a variety of decay channels. A recent computation suggests that $\mathcal{B}(Z^0 \to (c\bar{b}) + X) \approx 6 \times 10^{-4}$~\cite{Liao:2015vqa}.

The largest existing $e^+ e^- \to Z^0 \to \hbox{hadrons}$ data sets were recorded by experiments at CERN's Large Electron--Positron collider (LEP) during the 1990s. In samples of ($3.02,~3.9,\hbox{ and }4.2$) million hadronic $Z^0$ decays, the DELPHI, ALEPH, and OPAL Collaborations~\cite{Abreu:1996nz,*Barate:1997kk,*Ackerstaff:1998zf}  found a small number of candidates for the decays $B_c \to \jpsi \pi^+, \jpsi \ell^+ \nu, \hbox{and }\jpsi 3\pi$. Those few specimens were not sufficient to establish a discovery, but the experiments were able to bound combinations of branching fractions $\mathcal{B}$ as
\begin{equation}
\frac{\mathcal{B}(Z^0 \to B_c + X)}{\mathcal{B}(Z^0 \to \hbox{hadrons})}\,\mathcal{B}(B_c \to \left\{\begin{array}{l} \jpsi \pi^+ \\ \jpsi \ell^+\nu \\ \jpsi 3\pi \end{array}\right\}) \lesssim \hbox{few} \times 10^{-4},
\label{eq:brlims}
\end{equation}
at 90\% confidence level, where $X$ denotes anything. The relative simplicity of $e^+ e^- \to Z^0$ events and the boosted kinematics of resulting $B_c$ mesons suggest that a Tera-$Z$ factory might be a felicitous choice to investigate $\spec{2}{}{P}{} \to \spec{1}{}{S}{} + \gamma$ lines.

\section{Conclusions and outlook\label{sec:conclude}}
In this article, we have presented a new analysis of the spectrum of mesons with beauty and charm. First, we modified the traditional Coulomb-plus-linear form of the quarkonium potential to incorporate running of the strong coupling constant \alphas\ that saturates at a fixed value at long distances. The new frozen-\alphas\ potential incorporates both perturbative and nonperturbative aspects of quantum chromodynamics. Second, we have set aside the perturbative treatment of spin splittings, instead incorporating lessons from Lattice QCD and observations of the $(c\bar{c})$ and $(b\bar{b})$ spectra.

We look forward to additional experimental progress, first by confirming and elaborating the characteristics of the \spec{2}{}{S}{} levels reported by the CMS Collaboration~\cite{Sirunyan:2019osb}. Key observables are the mass of the \spec{2}{1}{S}{0} state, the splitting between the two lines, and the ratio of peak heights corrected for efficiencies and acceptance. It is also of interest to test whether the dipion mass spectra in the cascade decays $B_c^\prime \to B_c\pi^+\pi^-$ and $B_c^{*\prime} \to B_c^*\pi^+\pi^-$ follow the pattern seen in $\psi(2S) \to \jpsi\pi^+\pi^-$ and $\Upsilon(2S) \to \Upsilon(1S)\pi^+\pi^-$ decays. Although we expect the \spec{3}{}{S}{} levels to lie above flavor threshold, exploring the $B_c\pi^+\pi^-$ mass spectrum up through $7200\mev$ might yield indications of $\spec{3}{1}{S}{0} \to B_c\pi^+\pi^-$ and $\spec{3}{3}{S}{1} \to B_c^*\pi^+\pi^-$ lines. The presence of one or the other of these could signal interactions of bound states with open channels. Prospecting for narrow $B^{(*)}D^{(*)}$ peaks near threshold could yield evidence of $B_c$ states beyond the \spec{2}{}{S}{} levels.

The next frontier is the search for radiative transitions among $(c\bar{b})$ levels. The most promising candidate for first light is the  $\spec{2}{3}{P}{2}(6750) \to B_c^*\gamma$ transition. 
Determining the $B_c^*$ mass, perhaps by reconstructing $B_c^* \to B_c\gamma$, would provide an important check on lattice QCD calculations and a key input to future calculations.

Detecting the $B_c \to \tau\nu_\tau$ and $B_c \to p\bar{p}\pi^+$ decays would be impressive experimental feats, and would provide another test of the short-distance behavior of the ground-state wave function, complementing what will be learned from the $B_c^*$--$B_c$ splitting.

\appendix* {\label{append}}
\section{Strong coupling evolution\label{app:frozen}}
To make calculations with the frozen-\alphas\ potential, one must combine a linear term with a Coulomb term, $-4\alphas(r)/3r$, for which $\alphas(r)$ is characterized by the solid red curve of Figure~\ref{fig:alphar}. We present in Table~\ref{tab:alphasvals} numerical values of the strong coupling over the relevant range of distances, $0 \le r \le 0.8\fm$. The entries advance in steps of $\delta\ln r =0.1$.
{\begingroup
\squeezetable
\begin{table}[h]
\caption{Evolution of the strong coupling. \label{tab:alphasvals}}
 \begin{tabular}{@{\hskip 1em}c@{\hskip 4em}d@{\hskip -2em}}
 \toprule
 \multicolumn{1}{l}{\phantom{M}$r\text{ [fm]}$} & \multicolumn{1}{l}{$\alphas(r)$\B}\\
\colrule
$	0.0080	$   &   $	0.1706	$\\
$	0.0088	$   &   $	0.1742	$\\
$	0.0097	$   &   $	0.1780	$\\
$	0.0108	$   &   $	0.1819	$\\
$	0.0119	$   &   $	0.1862	$\\
$	0.0132	$   &   $	0.1908	$\\
$	0.0145	$   &   $	0.1957	$\\
$	0.0161	$   &   $	0.2007	$\\
$	0.0178	$   &   $	0.2061	$\\
$	0.0196	$   &   $	0.2116	$\\
$	0.0217	$   &   $	0.2174	$\\
$	0.0240	$   &   $	0.2235	$\\
$	0.0265	$   &   $	0.2299	$\\
$	0.0293	$   &   $	0.2365	$\\
$	0.0323	$   &   $	0.2434	$\\
$	0.0358	$   &   $	0.2505	$\\
$	0.0395	$   &   $	0.2579	$\\
$	0.0437	$   &   $	0.2659	$\\
$	0.0483	$   &   $	0.2743	$\\
$	0.0533	$   &   $	0.2829	$\\
$	0.0589	$   &   $	0.2915	$\\
$	0.0651	$   &   $	0.3001	$\\
$	0.0720	$   &   $	0.3087	$\\
$	0.0796	$   &   $	0.3171	$\\
$	0.0879	$   &   $	0.3252	$\\
$	0.0972	$   &   $	0.3330	$\\
$	0.1074	$   &   $	0.3403	$\\
$	0.1187	$   &   $	0.3471	$\\
$	0.1312	$   &   $	0.3533	$\\
$	0.1450	$   &   $	0.3590	$\\
$	0.1602	$   &   $	0.3640	$\\
$	0.1771	$   &   $	0.3685	$\\
$	0.1957	$   &   $	0.3723	$\\
$	0.2163	$   &   $	0.3757	$\\
$	0.2390	$   &   $	0.3786	$\\
$	0.2642	$   &   $	0.3811	$\\
$	0.2920	$   &   $	0.3832	$\\
$	0.3227	$   &   $	0.3849	$\\
$	0.3566	$   &   $	0.3864	$\\
$	0.3941	$   &   $	0.3876	$\\
$	0.4355	$   &   $	0.3886	$\\
$	0.4813	$   &   $	0.3895	$\\
$	0.5320	$   &   $	0.3902	$\\
$	0.5879	$   &   $	0.3908	$\\
$	0.6497	$   &   $	0.3913	$\\
$	0.7181	$   &   $	0.3917	$\\
$	0.7936	$   &   $	0.3920	$\\
\botrule
\end{tabular}
\end{table}
\endgroup}


\section*{Addendum\label{addendum}}
In the discussion surrounding Figure 8 of the published version of this Article~\cite{Eichten:2019gig}, we highlighted the possibility that E1 electric-dipole transitions from the $2P \to 1S$ levels might offer an imminent opportunity to establish orbitally excited levels We pointed to the $\spec{2}{3}{P}{2}(6750) \to B_c^*\gamma$ line as an especially promising target for experiment, because of the favorable production cross section and 409-MeV photon energy.
We did not specifically comment of prospects for establishing the $3P$ states. This Addendum repairs that omission.

We show in a new Figure~\ref{fig:e1trans} cross sections $\times$ branching fractions for the spectrum of E1 photons in decays of the \spec{3}{}{P}{J}\ to \spec{1}{}{S}{}\ levels. (Since the $3S$ levels should lie above flavor threshold, we neglect feed-down from $3S \to 3P$ transitions. Cross sections for the physical $3P_1^{(\prime)}$ states are
appropriately weighted mixtures of the \spec{3}{3}{P}{1}\ and \spec{3}{1}{P}{1} cross sections.)  Although the yields are approximately four times smaller than those for the $2P \to 1S$ lines, the higher photon energies may be a decisive advantage for detection. The $\spec{3}{3}{P}{2}(7154) \to B_c^*\gamma(777\mev)$ line is a particularly attractive target for experiment. 

Experiments at the Large Hadron Collider have demonstrated the feasibility of E1 spectroscopy in the $(b\bar{b})$ family, discovering and characterizing $\chi_{b1}^{\prime\prime}$ and $\chi_{b2}^{\prime\prime}$~\cite{Aad:2011ih,*Aaij:2014caa,*Aaij:2014hla,*[{}][{. Note that these articles label states by the radial quantum number, $n - L$.}]Sirunyan:2018dff}. Observation of some $(c\bar{b})$ $P$-wave states should be possible with the data sets now in hand.


\begin{figure}[b] 
   \includegraphics[width=\columnwidth]{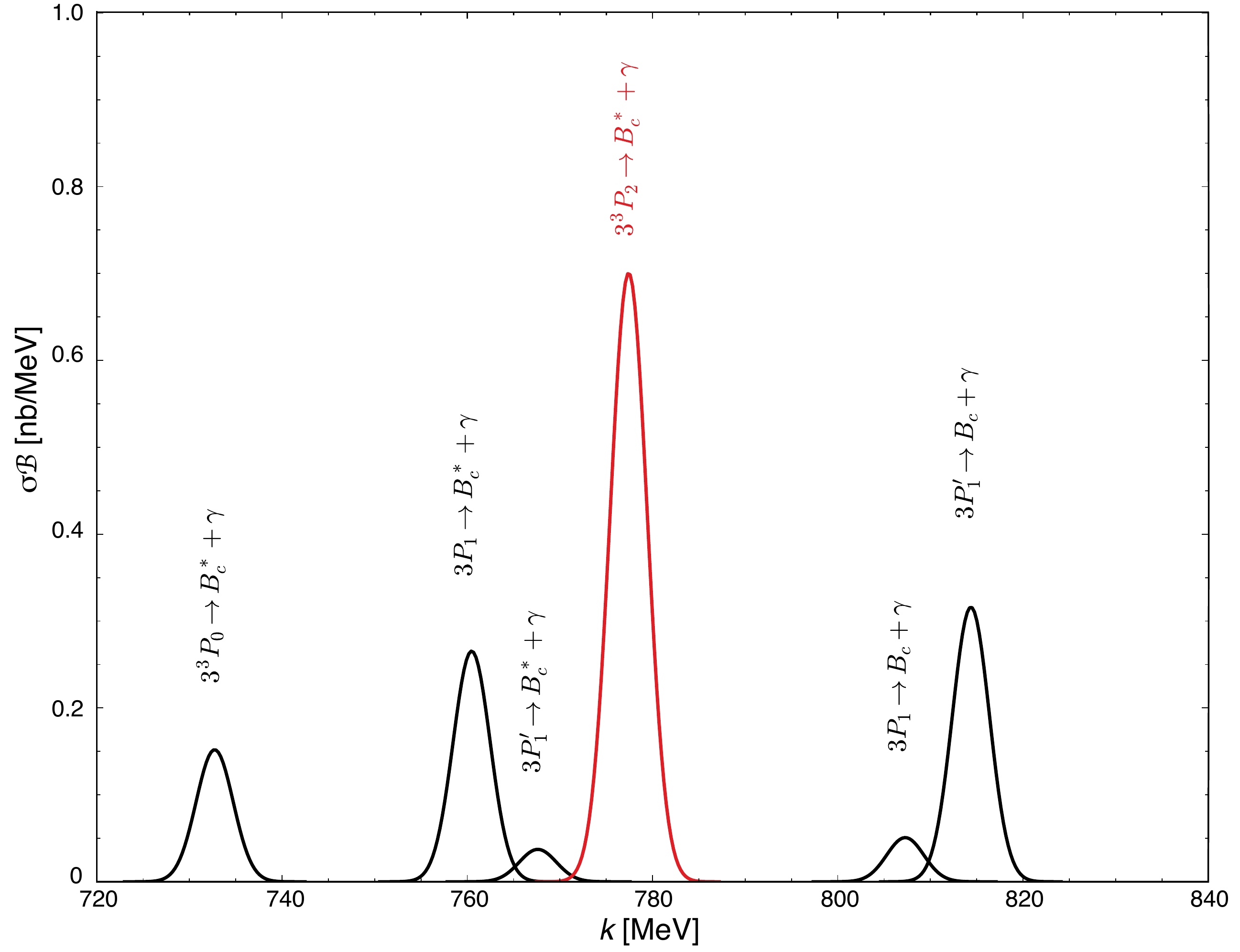} 
   \caption[]{Photon energies $k$ and predicted yields of  E1 transitions from $3P \to 1S$  $(c\bar b$) states. Photon momenta and E1 branching fractions are taken from from Table~VI; production rates are taken from Table~VIII. The  $3P$ masses inferred from transitions to $B_c^*$ will be shifted downward because of the  missing $B_c^* \to B_c \gammiss$ photon in the reconstruction. 
 We model Gaussian lineshapes with standard deviation $2\mev$.  \label{fig:e1trans}}
\end{figure}


\begin{acknowledgments}
This work was supported by Fermi Research Alliance, LLC under Contract No. DE-AC02-07CH11359 with the U.S. Department of Energy, Office of Science, Office of High Energy Physics.
\end{acknowledgments}

\bibliography{Bc2016plus}

\end{document}